\documentclass[submit]{elsarticle}

\usepackage[T1]{fontenc}
\usepackage{lmodern}
\usepackage{microtype}

\usepackage{algpseudocode}
\algnewcommand{\Whenever}[1]{\State \textbf{whenever} #1 \textbf{do}}
\algnewcommand{\EndWhenever}{}

\usepackage{float}
\usepackage{amsmath,amssymb,mathtools}
\usepackage{xcolor}

\usepackage{graphicx}
\usepackage{xcolor}
\usepackage{url}
\usepackage{booktabs}
\usepackage{multirow}
\usepackage{array}
\usepackage{longtable}
\usepackage{amsthm}
\newtheorem{definition}{Definition}
\usepackage{adjustbox}
\renewcommand{\arraystretch}{1.3}
\usepackage[justification=centering]{caption}
\usepackage{booktabs}
\usepackage{tabularx}
\usepackage{fontawesome5}
\usepackage{tabularx}
\usepackage{array}
\usepackage{longtable}

\usepackage{subcaption}

\usepackage{algorithm}
\usepackage{algpseudocode}

\usepackage{enumitem}
\usepackage{placeins}
\usepackage{tikz}
\usetikzlibrary{fit,trees,positioning,shapes,arrows,patterns,shapes.geometric,backgrounds,fadings,calc,arrows.meta}

\usepackage{smartdiagram}
\usesmartdiagramlibrary{additions}

\usepackage{pgfplots}
\usepackage{pgfplotstable}
\pgfplotsset{compat=1.18}
\usepgfplotslibrary{polar}
\usepackage{soul}
\usepackage{orcidlink}

\usepackage{forest}
\usepackage{rotating}
\usepackage{pifont}
\usepackage{tcolorbox}

\newcolumntype{L}[1]{>{\raggedright\arraybackslash}p{#1}}
\newcolumntype{Y}{>{\raggedright\arraybackslash}X}

\usepackage{caption}
\begin{document}

\begin{frontmatter}

\title{Dynamic Conflict Clustering with lightweight Learning and Strategic Repair for Distributed Constraint Satisfaction}

\author[inst1]{Raziyeh Moghaddas\orcidlink{0009-0001-7818-5362}}
\ead{raziyeh.moghaddas@swansea.ac.uk}

\author[inst1]{Fabio Caraffini\orcidlink{0000-0001-9199-7368}\corref{cor1}}
\ead{fabio.caraffini@swansea.ac.uk}

\affiliation[inst1]{
  organization={Department of Science and Engineering, Swansea University},
  city={Swansea},
  country={UK}
}

\journal{arXive}

\begin{abstract}Distributed Constraint Satisfaction Problems (DCSPs) model distributed decision-making in multi-agent systems, where agents assign values to local variables while satisfying shared constraints. However, many DCSP solvers still rely on broad, repeated peer-to-peer communication and are largely reactive, with limited support for strategic coordination or learning. This increases communication overhead, can lead to stagnation, and reduces the success of hard instances. Hence, we propose \emph{Dynamic Conflict Clustering-Incomplete} (DCC-Incomplete), a repair-based solver that combines conflict-centred decomposition, structured coordination, strategic repair, and lightweight learning for stagnation recovery. DCC-Incomplete groups the unsolved part of the problem into conflict clusters and restricts communication via a coordination layer to agents within the same conflict region. It also introduces memory-guided restart, enabling agents and clusters to pause, revise, and adapt repairs based on neighbouring conflict states. A thorough evaluation of DCC-Incomplete on distributed graph-colouring benchmarks with planted satisfying assignments shows that its memory-guided restart improves the success rate from 37.50\% to 61.11\%, reduces median runtime from 0.565s to 0.367s, and cuts median communication from 12.493 to 6.503 messages per constraint. These results suggest that dynamic conflict clustering, combined with lightweight strategic and learning-guided repair, improves both communication efficiency and robustness in incomplete DCSP solving.
\end{abstract}


\end{frontmatter}

\section{Introduction}
\label{sec:introduction}

Distributed Constraint Satisfaction Problems (DCSPs) provide a formal framework for modelling distributed decision-making tasks in which multiple agents control local variables and must collectively satisfy constraints involving one another \cite{yokoo1998distributed}. This formulation is well-suited to Multi-Agent System (MAS) settings, where problem information, decision authority, and computational responsibility are distributed among autonomous agents \citep{yokoo1998distributed,wooldridge2009introduction}. 
Historically, DCSP and related distributed constraint reasoning models have been applied to several decentralised coordination tasks in sensor networks \citep{zhang2005distributedStochastic}, low-power embedded devices for environmental sensing and surveillance \citep{farinelli2008maxsum,rogers2011bounded}, scheduling \citep{bartak2010constraint, moghaddas2025train}, and other engineering domains \citep{gelle2000constraint}. Recent impactful applications include smart-home device scheduling, where appliances and sensors across multiple homes are coordinated as a multi-agent optimisation problem~\citep{kluegel2017realistic}; decentralised intersection traffic management using distributed constraint optimisation to coordinate vehicles~\citep{vu2018decentralised}; smart distribution-grid restoration and microgrid coordination~\citep{elsharafy2017backfeed}; and sensor-based dynamic environments where dynamic Distributed Constraint Optimisation Problems (DCOPs) are solved by sensor-actuator agents~\citep{barambones2021applicability}. Distributed constraint reasoning has also been used for scheduling, such as multi-agent train scheduling in single-track railways~\citep{moghaddas2025train}. Together, these applications highlight the need to solve constraint problems without a single central controller with full access to all variables, constraints, and decisions.

Since the formalisation of DCSPs -- we follow the well-established formulation in \citep{yokoo1998distributed} (Definition \ref{def:dcsp}) -- several families of distributed constraint-solving algorithms have been proposed. \textit{Complete} search methods guarantee a solution when one exists but often require substantial coordination and messaging. Examples include Asynchronous Backtracking (ABT) from the early DCSP formalisation \citep{yokoo1998distributed}, Asynchronous Weak-Commitment Search (AWC) \citep{yokoo1995awc}, and Asynchronous Partial Overlay (APO), which uses cooperative mediation over overlapping subproblems \cite{silver2005cooperative, ma2019lifelong, li2022constrained,mailler2006apo}. Related  DCOP methods include ADOPT, an asynchronous complete optimisation algorithm \citep{modi2005adopt}, and DPOP, a dynamic-programming method based on pseudo-tree utility propagation \citep{petcu2005dpop}. In contrast, \textit{incomplete} and local-search methods trade completeness for scalability and responsiveness. Examples include the Distributed Stochastic Algorithm (DSA) \citep{zhang2005distributedStochastic}, the Maximum Gain Message algorithm (MGM) \citep{maheswaran2004mgm}, Distributed Breakout Algorithms (DBA) \citep{hirayama2005breakout}, Max-Sum message passing \citep{farinelli2008maxsum}, and the Generalized Distributed Breakout Algorithm (GDBA) \citep{okamoto2016gdba}. These methods advance distributed constraint reasoning but also show that communication design and agent behaviour remain central to solver performance.

\begin{definition}[Distributed Constraint Satisfaction Problem]
\label{def:dcsp}
A DCSP is defined as the tuple $\mathrm{DCSP}=\langle A, X, D, C, \mathrm{own} \rangle$, where:
\begin{itemize}
    \item $A=\{a_1,\ldots,a_n\}$ is a finite set of agents;
    \item $X=\{x_1,\ldots,x_n\}$ is a finite set of variables;
    \item $D=\{D_1,\ldots,D_n\}$ is the set of finite domains, where 
    $x_i \in D_i$;
    \item $C=\{c_1,\ldots,c_m\}$ is a finite set of constraints;
    \item $\mathrm{own}:X\rightarrow A$ maps each variable to the agent that owns it.
\end{itemize}
\end{definition}

Technically, many distributed constraint solvers use fixed interaction structures, neighbour-to-neighbour communication, or repeated message exchange over the constraint graph. In incomplete constraint solving, repair-based methods start from a complete assignment and iteratively modify (repair) conflicted variable values to reduce violations ~\citep{minton1992minimizing,yokoo2000review}. This repair requires communication to maintain consistency and coordinate decisions between related agents, which can become costly when only a small part of the problem is in conflict. Often, locally consistent agents still engage in checking, messaging, or synchronisation even though the difficulty is confined to a smaller conflict region. This motivates a conflict-centred approach, where computation and communication focus on the currently unsatisfied part of the problem instead of repeatedly activating the whole graph.

From a MAS-based perspective, many solvers are strong in reactive local repair and distributed communication, but provide limited support for strategic behaviour and 
learning during search. This is problematic because agents that only address immediate local conflicts may repeat similar repairs, fail to adapt to changes in neighbouring regions, or continue repair attempts that no longer improve the global 
solving process. As a result, limited strategic behaviour and learning can contribute to stagnation, unnecessary communication, and lower success rates in 
difficult instances.

This work addresses both the technical and behavioural gaps identified above. Technically, it reduces unnecessary communication by dynamically organising the unsolved part of the problem into local conflict-centred structures (referred to as conflict clusters) and by regulating information exchange through a coordination layer. Conflict clusters allow restricting repair activity to agents currently involved in conflicts, while the coordination layer improves communication between agents, leaders, and neighbouring clusters. Behaviourally, the proposed model goes beyond immediate local repair by adding strategic coordination mechanisms and a lightweight learning-guided restart component. Strategic behaviour is used in a broad MAS sense, inspired by the game-theoretic idea that an agent’s decision may depend on the possible actions or states of other decision-makers \citep{nash1951noncooperative,shoham2008multiagent}. Note that this is not a formal game-theoretic equilibrium model; instead, it applies strategic coordination practically by delaying, reconsidering, coordinating, and adapting repair decisions according to the state of neighbouring conflict regions. The learning component follows the principle of memory-based search, using information on recently visited or unproductive states to guide exploration \citep{glover1989tabu}. Here, the memory mechanism is lightweight and local to stagnation recovery; it is not a full Tabu Search (TS) algorithm.

To summarise, we propose \emph{Dynamic Conflict Clustering-Incomplete} (DCC-Incomplete), a repair-based solver for DCSPs that classifies agents according to their current conflict involvement, forms dynamic conflict clusters over the active conflict regions, and regulates communication through a dedicated coordination layer. This structure keeps repair local while enabling coordination across clusters when needed. The model also adds strategic behaviours--postponed proposal handling, notification-based coordination, and conditional cluster merging--and a memory-guided restart mechanism that improves stagnation recovery by avoiding repeated exploration of recently unproductive states.

The main contributions of this work are as follows:
\begin{itemize}
    \item It proposes a dynamic conflict-centred decomposition mechanism for incomplete DCSP solving, in which repair is focused on currently violated regions of the constraint graph.
    \item Introduces the novel construct of conflict clusters and an additional coordination layer designed to systematically structure inter-agent communication and to minimize superfluous interactions among agents that are not directly participating in the same conflict region.
    \item It incorporates strategic coordination behaviours, including postponed proposal handling, notification-based inter-cluster coordination, and conditional cluster merging.
    \item It adds a lightweight learning-guided restart mechanism for stagnation recovery, inspired by memory-based search.
\end{itemize}

The remainder of the article is organised as follows. Section~\ref{sec:proposed_solver} presents the proposed DCC-Incomplete solver and its coordination mechanisms. 
Section~\ref{sec:experimental_design} describes the benchmark design, the compared methods, and the evaluation metrics. Section~\ref{sec:experimental_results} presents and discusses the experimental results. Finally, Section~\ref{sec:conclusion} 
concludes the article and outlines future work.

\section{The proposed DCC-Incomplete solver}
\label{sec:proposed_solver}

\begin{figure}[ht]
    \centering
    \includegraphics[width=1\textwidth]{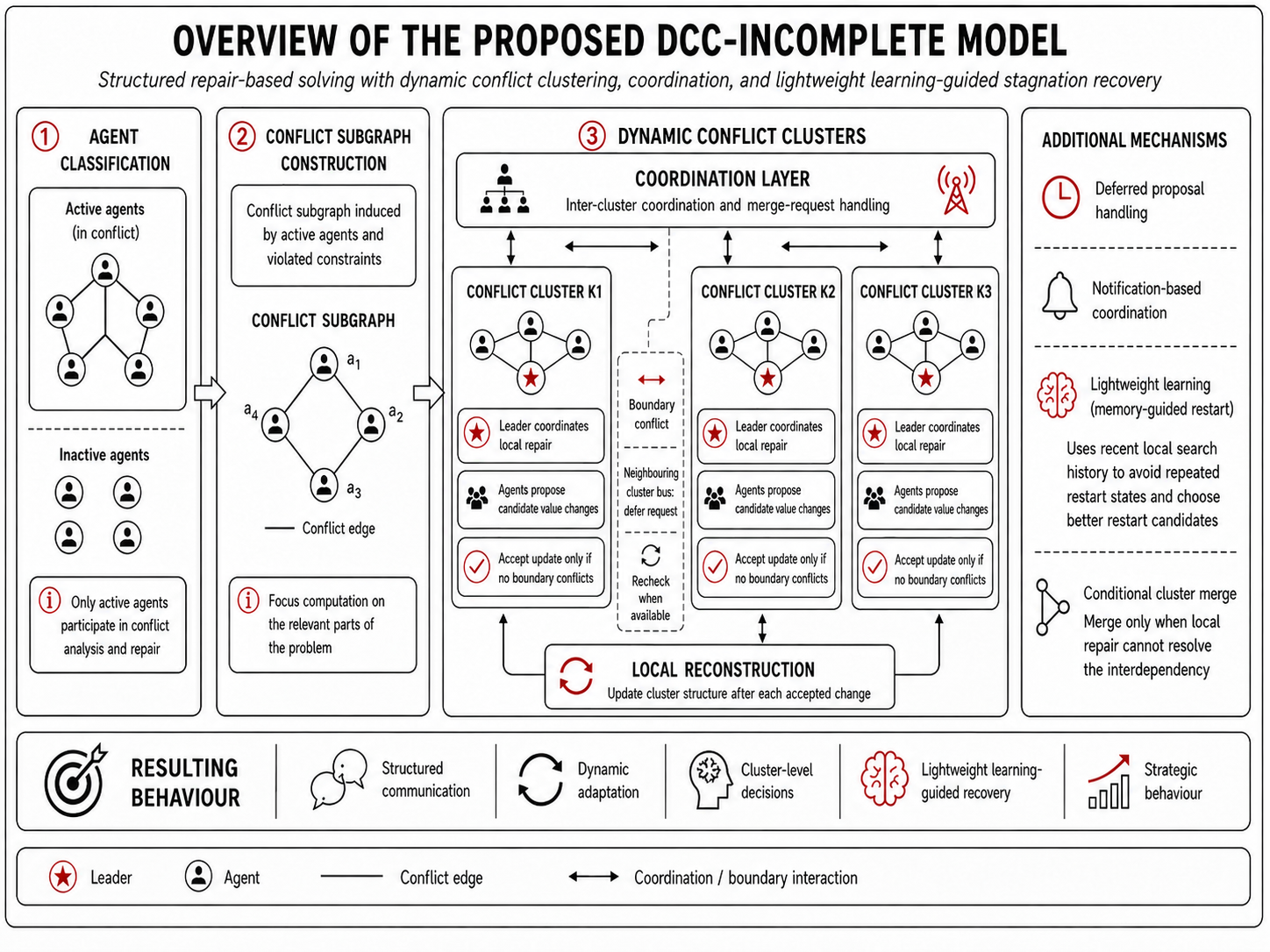}
    \caption{Architecture of DCC-Incomplete}
    \label{fig:Overview_of_the_Model}
\end{figure}

DCC-Incomplete is designed as an efficient repair-based solver for DCSPs. It is based on three key principles, namely dynamic conflict clustering, structured coordination, and localised repair mechanisms. Instead of repeatedly 
activating the entire constraint graph (Definition \ref{def:constraint_graph}), this solver focuses on the parts of the problem that are currently involved in conflicts. To achieve this, a preliminary module classifies agents, followed by the construction of a conflict subgraph, which ultimately enables the formation of dynamic conflict clusters governed by a dedicated coordination logic (as illustrated in Figure \ref{fig:Overview_of_the_Model}, blocks 1, 2, and 3).

\begin{definition}[Constraint Graph]
\label{def:constraint_graph}
Following Montanari’s constraint-network representation~\citep{montanari1974networks}, later standardised in constraint processing~\citep{dechter2003constraint} and adopted in DCS~\citep{yokoo1998distributed}, the constraint graph of a DCSP is an undirected graph $G=(V,E)$, where $V=X$ and $\exists c\in C \,\mid\, \{x_i,x_j\}\subseteq \mathrm{scp}(c),\, \forall (x_i,x_j)\in E$ (i.e., two variables are connected by an edge if they appear together in at least 
one constraint). 
\end{definition}

Note that the notation $\mathrm{scp}(c)\subseteq X$ is commonly used to denote the scope of a constraint $c$, that is, the subset of variables in $X$ that participate in, or are affected by, that constraint. For completeness, we recall for the non-specialist reader that a constraint graph-also commonly termed a constraint network-is the canonical graphical formalism used to model the interactions among variables that arise from a given set of constraints \citep{montanari1974networks,dechter2003constraint}. In DCS, the constraint graph provides the structural basis for identifying neighbouring agents (Definition \ref{def:neighbour}), local constraints (Definition \ref{def:localC}), and possible communication links \citep{yokoo1998distributed}. 

\begin{definition}[Agent neighborhood]\label{def:neighbour}
The neighbours of an agent $a_i$ in the constraint graph are denoted by the set $\mathcal{N}(i)=\{a_{j\neq i} \in A \mid (x_i,x_j)\in E\}$.
\end{definition}
\begin{definition}[Local constraints]\label{def:localC}
The set of local constraints of an agent $a_i$ is defined as $C_i=\{c\in C \mid x_i\in \mathrm{scp}(c)\}$.
\end{definition}
Furthermore, observe that, as in the most common setting, also in our framework, each agent is associated with exactly one variable, that is, $\mathrm{own}(x_i) = a_i$. The terms agent and variable are therefore nearly interchangeable. However, conceptually, a generic assignment $\alpha=\{x_1=v_1,\ldots,x_n=v_n\}$, where $v_i\in D_i$, is performed by the agent and contains variable values. A constraint $c\in C$ is satisfied under an assignment $\alpha$ 
if the values assigned to the variables in its scope meet the relation defining $c$. This is commonly denoted by $\alpha \models c$.  A DCSP is solved when all constraints are satisfied: $
\alpha \models c\,,\forall c\in C$.

\subsection{Agent classification}\label{sec:agentclass}

Agents are classified as \textit{active} or \textit{passive} based on their current involvement in conflicts. At a generic iteration \(t\), we define active agents as those that participate in at least one violated constraint. In contrast, passive agents are locally consistent with the current assignment and are therefore excluded from the repair process unless they subsequently become involved in a constraint violation. We can therefore partition the set of agents as \(A = A_A(t) \cup A_P(t)\), with \(A_A(t) \cap A_P(t) = \varnothing\), where the subscripts \(A\) and \(P\) denote the subsets of active and passive agents, respectively (Definition \ref{def:active_passive_agents}).

We define the ``local conflict cost'' of an agent in Definition \ref{def:local_cost} as a measure to identify which agents are 
currently involved in violated constraints and should therefore participate in 
repair.
\begin{definition}[Local Conflict Cost]
\label{def:local_cost}
Given an assignment $\alpha$ and a candidate value $v\in D_i$ for the variable $x_i$ owned by agent $a_i$, the \emph{local conflict cost} is
\[
\mathrm{LCost}_i(v)
=
\bigl|\{\,c\in C_i \mid c \text{ is violated under } \alpha[x_i\leftarrow v]\,\}\bigr|.
\]
\end{definition}

This, in turn, enables the introduction of the score associated with a given agent (Definition \ref{def:score}), which provides a more operational characterisation of active and passive agents (Definition \ref{def:active_passive_agents}).

\begin{definition}[Agent Score]
\label{def:score}
The score of agent $a_i$ under its current value $v_i$ is defined as
\[
\mathrm{Score}(a_i)=-\mathrm{LCost}_i(v_i).
\]
\end{definition}

Note that a score of $0$ indicates that the current value of $x_i$ violates no local constraint, whereas a negative score indicates participation in at least one violated constraint. Thus, $\mathrm{Score}(a_i)$ serves as a conflict indicator: values closer to $0$ correspond to lower local conflict.

\begin{definition}[Active ($A_A(t)$) and Passive ($A_P(t)$) Agent Sets]\label{def:active_passive_agents}
At iteration $t$, $A$ can be partitioned as follows:
\[
A_A(t)=\{a\in A \mid \mathrm{Score}_t(a)<0\},\quad
A_P(t)=\{a\in A \mid \mathrm{Score}_t(a)=0\}.
\]
\end{definition}

Classification is carried out dynamically in every iteration, and the active-agent partition serves as the basis for constructing the conflict subgraph.

\subsection{Conflict Subgraph Construction}
\label{subsec:conflict_subgraph}

DCC-Incomplete constructs a dynamic, conflict-focused representation based on the current assignment. This is regenerated whenever the assignment is updated, enabling the solver to concentrate its computation and communication on the part of the problem that is currently unsatisfied.

While the constraint graph in Definition \ref{def:constraint_graph} represents all potential variable interactions, we define a conflict subgraph (Definition \ref{def:conflict_subgraph}) that changes dynamically during the solving process to represent only currently violated interactions. 

\begin{definition}[Conflict Subgraph]
\label{def:conflict_subgraph}
Let $G=(V,E)$ be the constraint graph of a DCSP. The conflict 
subgraph at iteration $t$ is defined as $G_t^{-}=(V_t^{-},E_t^{-})$ where
$V_t^{-}=\{x_i\in X \mid a_i\in A_A(t)\}$  and $(x_i,x_j)\in E_t^{-}
\, \Longleftrightarrow \,
(x_i,x_j)\in E$ and at least one constraint involving $x_i$ and $x_j$ is violated under $\alpha_t$.
\end{definition}

\subsection{Dynamic Conflict Clusters}\label{subsec:dynamic_conflict_structures}

During each iteration \(t\), conflict subgraphs are processed to extract connected components through the $\mathrm{CC}(\dots)$ operator (Definition \ref{def:Connected_Component}) to obtain conflict clusters (Definition \ref{def:conflict_cluster}), and identify their respective leader agents. These leader agents (Definition \ref{def:cluster_leader}) coordinate the organisation of repair requests, receive candidate repair proposals, and evaluate potential boundary effects. 

\begin{definition}[Connected component]
\label{def:Connected_Component}
A connected component of an undirected graph $G=(V,E)$ is a maximal subgraph in which every pair of vertices is connected by a path.
\end{definition}
 
\begin{definition}[Conflict cluster]
\label{def:conflict_cluster}
A conflict cluster $\mathcal{K}_t$ of a given conflict subgraph $G_t^{-}$ in iteration $t$ of the DCC-Incomplete algorithm is defined as the set of all connected components $K_t^i$ in $G_t^{-}$, i.e., $\mathcal{K}_t=\mathrm{CC}(G_t^{-})=\{K_t^1,K_t^2,\ldots,K_t^{r_t}\}$.
\end{definition}

\begin{definition}[Cluster Leader]
\label{def:cluster_leader}
Let $\mathrm{Members}_t^k$ denote the agents whose variables belong to the conflict cluster $K_t^k$. The leader of $K_t^k$ is selected as the least conflicted member of the cluster: $
\mathrm{Leader}_t^k
=
\arg\max_{a_i\in \mathrm{Members}_t^k}
\bigl(\mathrm{Score}_t(a_i),-i\bigr)$\footnote{The first argument selects the agent whose score is closest to $0$, whereas the 
second argument is employed as a tie-breaking criterion, selecting the agent with the smallest identifier among those candidates.}.
\end{definition}

Note that each conflict cluster is a subgraph $K_t^k=(V_t^k,E_t^k)$
where $V_t^k\subseteq V_t^{-}$ and $,E_t^k\subseteq E_t^{-}$. The conflict clusters are disjoint and cover the conflict subgraph: $V_t^{-}=\bigcup_{k=1}^{r_t}V_t^k$, $V_t^i\cap V_t^j=\varnothing$ for  $i\neq j$. They correspond to dynamically identified, currently unresolved subproblems and may emerge, disappear, split, or merge as the underlying assignment evolves. This behaviour constitutes the foundation of the dynamic decomposition mechanism employed by DCC-Incomplete. 

We highlight that the repair process is directed toward the most conflicted agent in the cluster, which we refer to as the target agent (Definition \ref{def:target_agent}).

\begin{definition}[Target Agent]
\label{def:target_agent}
The target agent of a conflict cluster $K_t^k$ is define as
$
\mathrm{Target}_t^k
=
\arg\min_{a_i\in \mathrm{Members}_t^k}
\bigl(\mathrm{Score}_t(a_i), i\bigr).
$.
\end{definition}

Note that in the case of ties, the formula for selecting the target agent gives preference to the agent with the smallest identifier. Finally, the solver employs a cluster-level conflict metric, i.e., the cluster conflict load (Definition \ref{def:cluster_conflict_load}), to assess whether a given cluster still necessitates repair.

\begin{definition}[Cluster Conflict Load]
\label{def:cluster_conflict_load}
The conflict load of a conflict cluster $K_t^k$ is defined as $\mathrm{Load}_t^k=-\sum_{a_i\in \mathrm{Members}_t^k}\mathrm{Score}_t(a_i)$.
\end{definition}

It follows that, if 
$\mathrm{Load}_t^k = 0$, 
the corresponding cluster is considered locally resolved. Conversely, if 
$\mathrm{Load}_t^k > 0$, 
further repair operations are still necessary.

\subsection{Coordination layer}
\label{subsec:coordination_layer}

The purpose of the coordination layer (Definition \ref{def:coordination_layer}) is to manage repair and communication at the conflict-cluster level without centralising the solution process. It is a coordination record that stores for each conflict cluster $k\in\mathcal{K}_t$ its cluster membership, selected leader, local context, boundary information, conflict load, coordination state, deferred proposals, and stagnation information. These pieces of information are used to give leaders details to coordinate local repairs, monitor boundary conflicts, and manage inter-group dependencies.

\begin{definition}[coordination layer]
\label{def:coordination_layer}
The coordination layer of a generic $k^{th}$ conflict cluster $K_t^k$ is a record structured as follows:
\[
\resizebox{0.98\linewidth}{!}{$
\mathcal{L}_t(K_t^k)
=
\left\langle
K_t^k,\,
\mathrm{Members}_t^k,\,
\mathrm{Leader}_t^k,\,
\mathrm{Context}_t^k,\,
\mathrm{Boundary}_t^k,\,
\mathrm{Load}_t^k,\,
\sigma_t^k,\,
Q_t^k,\,
\mathrm{NoImproveCount}_t^k
\right\rangle
$}
\]
where
\begin{itemize}
    \item $\mathrm{Members}_t^k$ is the set of agents whose variables belong to $K_t^k$.    
    \item $\mathrm{Context}_t^k$ stores the current values needed to evaluate local repair moves, including cluster member values and relevant boundary neighbours (see Section \ref{subsubsec:boundary_conflict_handling} for details on boundaries).
    \item $\mathrm{Boundary}_t^k= \{(x_i,x_j)\in E \mid x_i\in V_t^k,\; x_j\notin V_t^k\}$ is the set of constraint-graph edges connecting the cluster to variables outside the cluster. 
    \item $\sigma_t^k \in \{\texttt{idle},\texttt{repairing},\texttt{waiting}\}$ Indicates the cluster’s current coordination state and whether it is available for repair or involved in a coordination dependency.
    \item $Q_t^k$ is the deferred-proposal queue for the cluster.
    \item $\mathrm{NoImproveCount}_t^k$ is a record of consecutive repair attempts that fail to improve the cluster state.
\end{itemize}
\end{definition}

The \texttt{idle} state means the cluster is available for repair. The \texttt{repairing} state means a repair action is in progress. The \texttt{waiting} state means the cluster cannot currently improve and is waiting for a neighbouring cluster to change before reconsidering deferred proposals.

The proposals in $Q_t^k$ are those that could improve the local cluster state but would create or maintain a boundary conflict with a neighbouring cluster. These proposals are postponed and may be reconsidered after a relevant boundary update. When a change is accepted in $K_t^k$, $Q_t^k$ is cleared ($Q_t^k \leftarrow \varnothing$), so outdated proposals are not reconsidered after the cluster context changes.

In conclusion, the coordination layer has two main functions. First, it minimises unnecessary communication by restricting fine-grained repair coordination to the relevant conflict cluster and its boundary. Second, it supports strategic coordination by allowing clusters to defer decisions, wait for changes in adjacent clusters, and trigger merging only when local repair proves insufficient.

\subsection{Communication Model}
\label{subsec:communication_model}

The purpose of the communication model is to avoid unrestricted peer-to-peer interaction while still allowing sufficient information exchange for local repair and boundary coordination. Communication is therefore limited to four channels: agent-to-leader communication, leader-to-agent communication, leader-to-leader communication, and internal coordination layer updates.

\subsubsection{Agent-to-Leader Communication} \label{subsubsec:Agent-to-Leader_Communication}

Agent-to-leader communication occurs when agents report current values, propose repairs, or indicate no improving move. These messages let the leader update the coordination layer record and choose the next repair action for the cluster.

The main message type is $\mathrm{VALUE\_UPDATE}(a_i,x_i,v_i)$, which reports agent $a_i$’s current variable value and updates the cluster context: $\mathrm{Context}_t^k[x_i]\leftarrow v_i$.

The second message type, $\mathrm{REPAIR\_PROPOSAL}(a_i,x_i,v_i^*,\mathrm{LCost}_i(v_i^*))$, informs the leader that agent $a_i$ proposes changing $x_i$ to $v_i^*$ with the corresponding local conflict cost.

The third message type, $\mathrm{NO\_IMPROVEMENT}(a_i)$, indicates that agent $a_i$ cannot find any improving local move in the current context.

\subsubsection{Leader-to-Agent Communication} \label{subsubsec:Leader-to-Agent_Communication}

Leader-to-agent communication initiates local repair and informs agents whether their proposals are accepted, postponed, or rejected for the current step. The leader does not compute the full solution centrally; it only coordinates the repair using information in the coordination layer.

The main repair request is $\mathrm{REPAIR\_REQUEST}(a_i,\mathrm{ContextSubset})$, where $\mathrm{ContextSubset}$ contains only the values agent $a_i$ needs to evaluate candidates.

If a proposal improves the cluster state and creates no boundary conflict, the leader sends $\mathrm{ACCEPT}(a_i,x_i,v_i^*)$, and the agent commits the value change.

If a proposal reduces the internal conflict load of the current conflict cluster, but causes or maintains a boundary conflict, the leader agent sends the message $\mathrm{DEFER}(a_i,x_i,v_i^*,\mathrm{LCost}_i(v_i^*))$.
The proposal is stored in the deferred queue $Q_t^k$ and may be reconsidered after a relevant boundary update.

\subsubsection{Leader-to-Leader Communication}

Leader-to-leader communication is limited to coordinating boundaries between neighbouring conflict clusters and is used only when a repair decision in one cluster may affect another via a boundary constraint.

The main boundary update message, $\mathrm{BOUNDARY\_UPDATE}(K_t^k,x_i,v_i)$, notifies neighbouring clusters of a changed boundary variable, and the receiving leader updates the corresponding boundary entry in its cluster context.

If a boundary conflict cannot be resolved locally, then the leader sends $\mathrm{MERGE\_REQUEST}(K_t^k,K_t^h)$, where $K_t^h$ is the neighbouring cluster in conflict. The neighbour replies with $\mathrm{MERGE\_RESPONSE}(K_t^h,K_t^k,\rho)$, where $\rho \in \{\texttt{ACCEPT},\texttt{POSTPONE}\}$.

A \texttt{POSTPONE} response means the neighbouring cluster still has internal repair options and should not merge yet. A \texttt{ACCEPT} response means local repair is insufficient and the clusters may be merged to resolve the dependency together.

For notification-based coordination, a leader can request updates when a neighbouring cluster changes state, $\mathrm{NOTIFY\_REQUEST}(K_t^k,K_t^h)$. After a relevant change, the neighbour sends $\mathrm{NOTIFY\_UPDATE}(K_t^h,K_t^k)$, allowing deferred proposals to be reconsidered under the updated boundary context.

\subsubsection{coordination layer Internal Updates}

The fourth channel is not a network message but logs internal updates to the coordination layer after repairs or coordination actions, including changes to $\mathrm{Context}_t^k,\ \mathrm{Boundary}_t^k,\ \mathrm{Load}_t^k,\ \sigma_t^k,\ Q_t^k,\ \mathrm{NoImproveCount}_t^k$.

This mechanism keeps the leader’s cluster context current for repair and blocks outdated proposals and stale boundary data from influencing the search.

Thus, the communication model fulfils DCC-Incomplete’s main design goal: repairs remain local when possible, and inter-cluster communication occurs only when boundary dependencies require coordination.

\subsection{Repair and Coordination Procedure}
\label{subsec:repair_coordination}

The repair procedure of DCC-Incomplete works at the conflict cluster level. Each conflict cluster is an unsatisfied region of the problem, and the coordination layer supplies the information needed to coordinate repair without centralising solving. The leader coordinates the repair cycle, while agents still evaluate candidate values in a distributed way. This preserves agent autonomy, avoids centralising each conflict cluster, and prevents the leader from becoming a computational bottleneck.

At each repair step, the leader selects as target the agent whose current value contributes most to the cluster's conflict load. The leader sends this agent a repair request and the relevant context from the coordination layer, including the current values of variables needed to evaluate candidate repairs and any required boundary-neighbour values.

The target agent evaluates its candidate values locally. For each $v\in D_i$, it computes the local conflict cost $\mathrm{LCost}_i(v)$ and returns a repair proposal to the leader. The leader does not compute all candidate values centrally; it only coordinates the decision by comparing the proposal with the current cluster state and checking whether it affects boundary consistency.

A proposal has three possible outcomes. First, if it improves the cluster state and does not create a boundary conflict, it is accepted and committed. Second, if it is locally useful but conflicts with a neighbouring conflict cluster, it is deferred and stored in its queue $Q_t^k$. Third, if no candidate value improves the state, the cluster records a no-improvement event, which supports stagnation detection and may later trigger the memory-guided restart mechanism in Section~\ref{subsec:memory_guided_restart}.

This process is implemented as in Algorithm \ref{alg:repair_cycle}.

\subsubsection{Boundary-Conflict Handling}
\label{subsubsec:boundary_conflict_handling}

Boundary conflicts arise when a candidate repair within a conflict cluster affects a constraint involving a variable outside that cluster. These cases matter because accepting the repair may improve the current cluster but worsen a neighbouring one. DCC-Incomplete therefore checks boundary consistency before committing to a proposed value change.

Let $K_t^k$ be the current conflict centre and let $(x_i,x_j)\in \mathrm{Boundary}_t^k$ be a boundary edge, where $x_i\in V_t^k$ and $x_j\notin V_t^k$. If changing $x_i$ to a candidate value $v_i^*$ would violate a boundary constraint involving $x_j$, the proposal is treated as boundary-conflicting. The leader then stores the proposal in $Q_t^k$ rather than accepting it immediately. This allows the solver to preserve a potentially useful move while avoiding premature decisions that may create unnecessary inter-cluster disruption.

Boundary-conflict handling constitutes one of the primary mechanisms through which the model achieves strategic coordination. Rather than responding solely to immediate, localised improvements, each cluster evaluates the impact of a prospective move on adjacent conflict regions. This facilitates more regulated information exchange and mitigates the occurrence of recurrent local adjustments that merely propagate conflicts across cluster boundaries.

\subsubsection{Deferred Proposal Handling}
\label{subsubsec:deferred_proposal_handling}

A deferred proposal is a candidate repair that might improve the current cluster but cannot be safely committed under the current boundary context. When a proposal is deferred, it enters $Q_t^k$, and the leader stores the candidate value, its local cost, and the boundary dependency that blocked immediate acceptance. 

Deferred proposals are reconsidered only when boundary context changes, usually after a neighbouring conflict cluster commits a value and sends an update. The leader then re-evaluates the proposal: if it no longer conflicts at the boundary and still improves the cluster state, it may be accepted; otherwise, it is discarded. When a cluster commits any accepted change, it clears its deferred queue ($Q_t^k \leftarrow \varnothing$) because the local context has changed and earlier proposals may no longer match the cluster’s current state.

\subsubsection{Postponed Merge-Request Handling}
\label{subsubsec:postponed_merge_request_handling}

When a boundary conflict cannot be resolved by local repair or by waiting for a deferred proposal, the current cluster may request a merge with a neighbouring cluster. However, if the neighbour still has internal repair options, the merge can be postponed. This avoids unnecessary cluster growth and lets each cluster resolve its own conflicts before merging.

A merge request is handled conditionally. If the neighbouring cluster is still improving, it replies \texttt{POSTPONE}, and the requesting cluster waits for a later update and may reconsider. If the neighbouring cluster cannot improve further, it replies \texttt{ACCEPT}, and the two clusters may merge to solve the shared dependency jointly. This mechanism enables controlled cooperation between clusters. They merge only when local repair and deferred coordination are insufficient, so DCC-Incomplete keeps repair local when possible and widens its scope only when necessary.

\subsubsection{Conditional Cluster Merging}
\label{subsubsec:conditional_cluster_merging}

Conditional cluster merging is used when two neighbouring conflict clusters are strongly coupled through unresolved boundary conflicts. When a merge is accepted, the clusters are replaced by a larger conflict cluster whose members are the union of the originals, and the conflict subgraph and coordination layer are rebuilt for the merged region.
Let $K_t^k$ and $K_t^h$ be neighbouring conflict clusters. If the merge is accepted, the new cluster is $K_t^{kh} = K_t^k \cup K_t^h$. The latter receives an updated member set, boundary set, leader, context, and conflict load. Repair then continues within the merged cluster.

Conditional merging lets the solver match the repair scope to the current conflict structure. If conflicts are local, clusters stay separate. If they span a boundary, the solver temporarily enlarges the scope so dependent variables are repaired together.

\subsubsection{Local Reconstruction}
\label{subsubsec:local_reconstruction}

After each accepted value change, DCC-Incomplete updates the local conflict state. This is necessary because a single repair action can alter active agents, violated constraints, and conflict-cluster connections. Instead of rebuilding the entire problem from scratch, the solver reconstructs only the affected conflict structures using the updated assignment.

The reconstruction step updates agent scores, active/passive agent sets, the conflict subgraph and clusters, cluster leaders, boundary sets, and coordination records. Solved clusters are removed. Repairs that create new conflicts can introduce new active agents or conflict edges. When the conflict structure changes, clusters may split, merge, or be replaced by new connected components.

Local reconstruction keeps the model dynamic by ensuring repair targets the current conflict state rather than an outdated view of the problem. 

This process is implemented as in Algorithm \ref{alg:local_reconstruction}. 

\subsubsection{Stagnation Detection}
\label{subsubsec:stagnation_detection}

DCC-Incomplete tracks stagnation (Definition \ref{def:stagnation}) using $\mathrm{NoImproveCount}_t^k$ in the coordination layer record. If a repair attempt does not improve the cluster state, the counter increases; when an accepted improvement is committed, the counter resets.

\begin{definition}[Stagnation]\label{def:stagnation}
A conflict cluster is stagnating when repeated\footnote{A fixed number of $\tau=2$ failures is commonly adopted for this parameter.} repair attempts fail to reduce its conflict load.  
\end{definition}

Stagnation detection is crucial because a repair-based incomplete solver may repeatedly revisit similar local states. If no improving move exists, continuing the same repair pattern wastes computation and communication. Thus, when $\mathrm{NoImproveCount}_t^k$ reaches a predefined threshold, the solver triggers a stagnation-recovery procedure. In the final DCC-Incomplete version, this is done via the lightweight learning method described in Section~\ref{subsec:memory_guided_restart}.

Overall, the repair procedure combines decentralised value assessment with structured, cluster-level coordination. Individual agents evaluate their own candidate values, while designated leaders manage proposal selection, boundary checks, deferred processing, merge requests, reconstruction, and stagnation monitoring. This division of responsibilities preserves the agents’ distributed behavior, reduces unnecessary communication, and enables more informed repair decisions.

\subsection{Memory-Guided Restart}
\label{subsec:memory_guided_restart}

A purely random restart can help the solver escape stagnation, but it may also restart the cluster into recently explored or unproductive local assignments. DCC-Incomplete overcomes this problem through a lightweight memory-guided restart mechanism.

Memory-guided restart does not perform an exhaustive memory-driven search like TS \citep{glover1989tabu}; instead, it borrows TS ideas to limit cycling and repeated moves in local search by avoiding recently ineffective assignments during stagnation recovery. Thus, it serves as a short-term, learning-guided restart mechanism, local to the affected conflict cluster.

In practice, each conflict cluster $K_t^k$ has a memory slot $M_t^k$ that stores recently visited or unproductive local assignments $\alpha_t^k
=
\{x_i=v_i \mid a_i\in \mathrm{Members}_t^k\}$. If the cluster fails to improve for a predefined number $\tau$ of repair attempts, its current local assignment is added to memory, $M_t^k \leftarrow M_t^k \cup \{\alpha_t^k\}$ ( $\tau=2$ in this study).

The restart is applied only to the stagnating conflict cluster, not the whole problem, keeping recovery local and avoiding changes to passive agents or progressing clusters. During restart, the solver generates a temporary pool of `\(p\)' candidate local assignments for variables in the stagnating cluster;\footnote{In the reported experiments, the candidate-pool size is fixed at \(p=25\).} these are restart alternatives, not a training dataset. It avoids candidate assignments already in the memory set whenever possible: $\alpha_{\mathrm{new}}^k \notin M_t^k$.Each candidate assignment is scored with a memory-aware rule: for assignment $x$,  
$S(x)=\lambda_1 I(x)+\lambda_2 B(x)+\lambda_3 R(x)-\lambda_4 D(x)$,  
where $I(x)$ is the number of internal violations in the conflict cluster, $B(x)$ the number of boundary violations with neighbouring clusters, $R(x)$ the repeated-state penalty, and $D(x)$ the number of variables changed from the previous local assignment. The parameters $\lambda_1,\lambda_2,\lambda_3,\lambda_4$ set their relative importance. Lower scores are preferred. The first two terms penalise candidates that preserve internal or boundary conflicts. The third term, calculated by comparing the candidate assignment $x$ with the restart memory $M_t^k$, discourages returning to recently explored unproductive local assignments. The fourth term rewards diversity by favouring candidates that move further from the stagnated configuration. The candidate with the lowest score becomes the new local assignment for the stagnating conflict cluster. 

After the restart, affected agents update their values, the coordination layer refreshes, and conflict structures rebuild. The restarted cluster’s no-improvement counter is reset ($\mathrm{NoImproveCount}_t^k \leftarrow 0$), and the deferred queue is cleared ($Q_t^k \leftarrow \varnothing$), because the local context has changed and previous proposals may be invalid.

 Algorithm \ref{alg:memory_guided_restart} describes the memory-guided restart procedure and provides a clear overview of the entire process.  Here, \(p\) denotes the candidate-pool size, i.e., the number of candidate local assignments generated during one restart.

\subsection{The DCC-Incomplete algorithm}
\label{subsec:overall_procedure}

 We report the overall architecture of the proposed framework in Algorithm~\ref{alg:overall_dcc}.

 \begin{algorithm}[ht!]
\caption{$\textsc{DCC-Incomplete}$}
\label{alg:overall_dcc}
\scriptsize
\begin{algorithmic}[1]

\State Fetch DCSP $\langle A,X,D,C,\mathrm{own}\rangle$ and user-specified  hyperparameters ($T_{max}=150$, $\tau=2$, $p=25$)
\State $\alpha \leftarrow \alpha_0$ \Comment{$\alpha_0$ is the initial assignment, usually a random guess}
\State $t \leftarrow 0$
\State $M \leftarrow \emptyset$ \Comment{Initialise memory records for stagnation recovery; Section~\ref{subsec:memory_guided_restart}}

\While{$t < T_{\max}$} \Comment{$T_{\max}$ is the pre-fixed max number of iteration}

    \State Compute $\mathrm{Score}(a),\,\forall a\in A$, and form $A_A(t)$ and $A_P(t)$ \Comment{Section \ref{sec:agentclass}}
    \If{there are no active agents}\Comment{i.e., $A_A(t) =\emptyset$}
        \State \Return $\alpha$ \Comment{All constraints are satisfied}
    \EndIf
    
    \State Generate $\mathcal{K}_t$ and build or update $\mathcal{L}_t(K_t^k),\,\forall K_t^k \in \mathcal{K}_t$\Comment{Section \ref{subsec:conflict_subgraph} -- \ref{subsec:coordination_layer}} 
    
    \ForAll{$K_t^k \in \mathcal{K}_t$}

        \State $\mathit{status} \leftarrow \textsc{RepairCycle}(K_t^k,\mathcal{L}_t(K_t^k),\tau,p)$\Comment{Algorithm \ref{alg:repair_cycle}}
        
        \If{$\mathit{status} \in \{\texttt{updated},\texttt{solved},\texttt{split},\texttt{merged}\}$}
            \State $(\mathcal{K}_t,\mathcal{L}_t) 
             \gets \textsc{LocalReconstruction}(K_t^k,\alpha,\mathcal{K}_t,\mathcal{L}_t)$ 
\Comment{Algorithm~\ref{alg:local_reconstruction}}
        \EndIf
    \EndFor
    
    \State Update $\alpha$ with accepted value changes \Comment{Section~\ref{subsec:repair_coordination}}
    \State $t \leftarrow t+1$
    
\EndWhile

\State \Return $\alpha$ \Comment{The final assignment reached; satisfying if solved} 
\end{algorithmic}
\end{algorithm}

From the overview in Algorithm~\ref{alg:overall_dcc}, it is straightforward to verify that once all agents become passive, the current assignment satisfies all constraints and the algorithm terminates successfully. Otherwise, due to its inherently incomplete nature, the solver returns the final assignment reached within the predefined iteration limit.

Algorithm~\ref{alg:repair_cycle} shows how DCC-Incomplete combines 
distributed local value evaluation with structured cluster-level coordination. The 
leader manages the repair cycle, deferred proposals, boundary dependencies, and merge 
requests, while agents remain responsible for evaluating their own candidate values.

\begin{algorithm}[ht!]
\caption{$\textsc{RepairCycle}(K_t^k,\mathcal{L}_t(K_t^k),\tau,p)$}
\label{alg:repair_cycle}
\scriptsize
\begin{algorithmic}[1]

\State Set $\sigma_t^k \leftarrow \texttt{repairing}$

\ForAll{$(a_i,x_i,v_i^*,K_t^h,\mathrm{LCost}_i(v_i^{*}))\in Q_t^k$} \Comment{Recheck deferred proposals (Section \ref{subsubsec:deferred_proposal_handling})}
    \If{$(a_i,v_i^*)$ no longer creates a boundary conflict with $K_t^h$}
        \State Accept the proposal and update the value of $a_i$
        \State$Q_t^k \leftarrow \varnothing$ \Comment{Clear the deferred queue}
        \State Reset $\mathrm{NoImproveCount}_t^k \leftarrow 0$
        \State \Return \texttt{updated} \Comment{the procedure break here ans return the value}
    \EndIf
\EndFor

\State Order $K_t^k$ members by conflict involvement \Comment{ Request repair from target agents}
\State Place the leader last in the target order

\ForAll{target agent $a_i$ in the ordered target list}
    \State Send $\mathrm{REPAIR\_REQUEST}$ to $a_i$ with the relevant context \Comment{ Section \ref{subsubsec:Leader-to-Agent_Communication}}
    \State Compute $\mathrm{LCost}_i$ for $a_i$ and returns a response \Comment{Section~\ref{subsubsec:Agent-to-Leader_Communication}}

    \If{$a_i$ returns $\mathrm{NO\_IMPROVEMENT}$}
        \State Continue to the next target agent 
    \EndIf

    \If{$a_i$ returns $\mathrm{REPAIR\_PROPOSAL}(a_i,v_i^*,\mathrm{cost})$}
        \If{$(a_i,v_i^*)$ creates no boundary conflict} \Comment{Section~\ref{subsec:repair_coordination}}
            \State Accept the proposal and update the value of $a_i$
            \State Clear the deferred queue: $Q_t^k \leftarrow \varnothing$
            \State Reset $\mathrm{NoImproveCount}_t^k \leftarrow 0$
            \State \Return \texttt{updated}\Comment{the procedure break here and return the value}
        \Else
            \State Identify the neighbouring conflict cluster $K_t^h$
            \State Store $(a_i,v_i^*,K_t^h)$ in $Q_t^k$
            \State Send $\mathrm{DEFER}$ to $a_i$
        \EndIf
    \EndIf
\EndFor

\If{$Q_t^k \neq \varnothing$} \Comment{Handle unresolved boundary dependencies (Section~\ref{subsubsec:postponed_merge_request_handling})}
    \State Select a deferred proposal $(a_i,v_i^*,K_t^h)$ from $Q_t^k$
    \State Send $\mathrm{MERGE\_REQUEST}(K_t^k,K_t^h)$ to the neighbouring cluster

    \If{the neighbouring cluster accepts the merge}
        \State Merge $K_t^k$ and $K_t^h$
        \State \Return \texttt{merged}
    \Else
        \State Keep the proposal deferred
        \State Set $\sigma_t^k \leftarrow \texttt{waiting}$
        \State \Return \texttt{waiting} \Comment{break, and exit the repair cycle}
    \EndIf
\EndIf

\State Increase $\mathrm{NoImproveCount}_t^k$\Comment{Stagnation detection; Definition~\ref{def:stagnation}}

\If{$\mathrm{NoImproveCount}_t^k \geq \tau$}
    \State $\textsc{MemoryGuidedRestart}(K_t^k,\mathcal{L}_t(K_t^k),M_t^k,p)$ \Comment{Algorithm~\ref{alg:memory_guided_restart}; Section~\ref{subsec:memory_guided_restart}}
    \State Clear the deferred queue: $Q_t^k \leftarrow \varnothing$
    \State Reset: $\mathrm{NoImproveCount}_t^k \leftarrow 0$
    \State \Return \texttt{updated}
\EndIf

\State Set $\sigma_t^k \leftarrow \texttt{idle}$
\State \Return \texttt{waiting}

\end{algorithmic}
\end{algorithm}

Algorithm \ref{alg:memory_guided_restart} implements the Memory-guided restart process. It is activated when a conflict cluster repeatedly fails to improve. 
The restart is local to the stagnating conflict cluster and does not directly change 
variables outside that cluster. Its purpose is to move the cluster away from recently 
unproductive local assignments while keeping the learning component lightweight.

Algorithm \ref{alg:local_reconstruction}, Local Reconstruction, updates the solver’s dynamic conflict structures after an accepted repair, merge, or restart. Using the current assignment, it recomputes local costs and agent scores, updates the active/passive classification, rebuilds the conflict subgraph, and extracts the new connected components. Depending on the resulting structure, the affected cluster may be removed if solved, refreshed if it remains as one component, or replaced by new ConflictClusters if it splits. This keeps the CoordinationLayer consistent with the current conflict state and prevents the solver from continuing with outdated cluster information.

\begin{algorithm}[ht!]
\caption{$\textsc{MemoryGuidedRestart}(K_t^k,\mathcal{L}_t(K_t^k),M_t^k,p)$}
\label{alg:memory_guided_restart}
\scriptsize
\begin{algorithmic}[1]

\State Let $\alpha_t^k$ be the current local assignment of $K_t^k$
\State Add $\alpha_t^k$ to memory: $M_t^k \leftarrow M_t^k \cup \{\alpha_t^k\}$  \Comment{Section~\ref{subsec:memory_guided_restart}}
\State Initialise candidate set $\mathcal{R} \leftarrow \varnothing$

\For{$r=1$ to $p$}
    \State Generate a candidate local assignment $\hat{\alpha}_r^k$ for the variables in $K_t^k$
    \If{$\hat{\alpha}_r^k \notin M_t^k$}
        \State Compute internal conflict cost $I(\hat{\alpha}_r^k)$
        \State Compute boundary conflict cost $B(\hat{\alpha}_r^k)$
        \State Compute repeated-state penalty $R(\hat{\alpha}_r^k)$
        \State Compute diversity value $D(\hat{\alpha}_r^k)$
        \State Compute candidate score:
        $S(\hat{\alpha}_r^k)
        =
        \lambda_1 I(\hat{\alpha}_r^k)
        +
        \lambda_2 B(\hat{\alpha}_r^k)
        +
        \lambda_3 R(\hat{\alpha}_r^k)
        -
        \lambda_4 D(\hat{\alpha}_r^k)$
        \State Add $\hat{\alpha}_r^k$ to $\mathcal{R}$
    \EndIf
\EndFor

\If{$\mathcal{R} \neq \varnothing$}
    \State Select the candidate assignment with the lowest score $S$
\Else
    \State Generate a random local assignment for $K_t^k$
\EndIf

\State Apply the selected assignment to the variables in $K_t^k$
\State Clear the deferred queue: $Q_t^k \leftarrow \varnothing$
\State Reset $\mathrm{NoImproveCount}_t^k \leftarrow 0$
\State \Return updated local assignment

\end{algorithmic}
\end{algorithm}

\begin{algorithm}[ht!]
\caption{$\textsc{LocalReconstruction}(K_t^k,\alpha_t,\mathcal{K}_t,\mathcal{L}_t)$ }
\label{alg:local_reconstruction}
\scriptsize
\begin{algorithmic}[1]

\ForAll{agents $a_i \in \mathrm{Members}_t^k$}\Comment{Recompute local scores} (Section~\ref{subsubsec:local_reconstruction})
    \State Use current assignment to recompute $\mathrm{LCost}_i$ and updated $\mathrm{Score}(a_i)$ 
\EndFor
\State Update $A_A(t)$ and $A_P(t)$\Comment{Reclassify agents} (Definition~\ref{def:active_passive_agents})
\State Rebuild $G_t^-$ and extract $\mathrm{CC}(G_t^-)$ \Comment{process new conflict subgraph (Definitions~\ref{def:conflict_subgraph}--\ref{def:conflict_cluster}) }

\If{no connected component remains} \Comment{Cluster solved; Section~\ref{subsubsec:local_reconstruction}}
    \State Remove $K_t^k$ from $\mathcal{K}_t$
    \State Remove $\mathcal{L}_t(K_t^k)$
    \State \Return $\langle\mathcal{K}_t,\mathcal{L}_t\rangle$ \Comment{Cluster solved; exit the reconstruction procedure}
\ElsIf{exactly one connected component remains}
    \State Update the membership of $K_t^k$
    \State Refresh its leader, target, boundary set, context, and conflict load
    \State Remove invalid deferred proposals from $Q_t^k$
    \State Set $\sigma_t^k \leftarrow \texttt{idle}$
    \State \Return $\langle\mathcal{K}_t,\mathcal{L}_t\rangle$ \Comment{Cluster updated; exit the reconstruction procedure}
\Else
    \State Remove old $K_t^k$ from $\mathcal{K}_t$
    \ForAll{new connected components}
        \State Create a new conflict cluster
        \State Select a leader and target
        \State Build its context, boundary set, conflict load, and $\mathcal{L}_t$ record
        \State Initialise $Q_t^k$ and $\mathrm{NoImproveCount}_t^k$
    \EndFor
    \State \Return $\langle\mathcal{K}_t,\mathcal{L}_t\rangle$ \Comment{Cluster split; exit the reconstruction procedure}
\EndIf

\end{algorithmic}
\end{algorithm}

The pseudocode presented in this section offers an operational overview of the systems previously described in the accompanying narrative. For the sake of reproducibility, we additionally provide the complete source code of the algorithms the entire experimental setup employed in this study in the associated repository \citep{bib:ZenodoDCC-Incomplete}.

\section{Experimental Design}
\label{sec:experimental_design}

We use a distributed graph colouring benchmark with planted satisfying assignments, which enables the evaluation of the solver on satisfiable DCSP instances while systematically controlling the graph size, constraint density, and initial conflict structure. We first perform a functionality assessment of the proposed algorithm and then we evaluate it, along with one of its variants, and conduct a comparative analysis against three well-established distributed constraint reasoning baselines.

\subsection{Benchmark Problem: Distributed Graph Colouring}
\label{subsec:benchmark_problem}

The benchmark problem in this study is graph colouring \citep{jensen1995graph}, where colours are assigned to vertices so that adjacent vertices receive different colours. This can be formulated as a DCSP, where each vertex is 
represented by one agent, and each agent controls the colour of its corresponding variable. Formally, let $G=(V,E)$ be an undirected graph with vertices $V=\{x_1,\ldots,x_n\}$, where each $x_i$ is a variable owned by agent $a_i$. Each variable has domain $D_i=\{1,\ldots,q\}$ of available colours. For every edge $(x_i,x_j)\in E$ there is a binary constraint $x_i \neq x_j$ requiring adjacent variables to have different colours. An assignment $\alpha$ is a valid colouring if $\alpha(x_i)\neq \alpha(x_j)$ for all $(x_i,x_j)\in E$.

Within the DCSP framework, a violated edge is equivalent to an unsatisfied binary constraint. This property facilitates the identification of conflict regions, as the connected components formed by violated edges inherently represent localised conflict structures. These structures can subsequently be exploited to analyse conflict cluster formation, perform repair operations, manage boundary coordination, and support reconstruction processes.

The benchmark instances are satisfiable distributed graph-colouring problems with a planted solution: a valid colouring is generated first, and the graph is then built so that this colouring remains valid. Thus, every instance is known to be satisfiable. An unsuccessful run only means the solver failed to find a satisfying assignment within the time limit, not that the instance is unsatisfiable.

We control three main factors: graph size, graph density, and random seed variation. Graph size is set by the number of variables $n$, density by four graph families, and random variation by the initial-assignment seeds (see Table \ref{tab:benchmark_families} for details).

\begin{table}[ht!]
\centering
\caption{Benchmark graph families used in the evaluation.}
\label{tab:benchmark_families}
\begin{adjustbox}{max width=\textwidth}
\begin{tabular}{llll}
\hline
 \textbf{Family (F)} & \textbf{Edge factor} & \textbf{Graph Size ($n$)} & \textbf{Role} \\
\hline
 Sparse (S)                   & 2.0 & $20, 50, 100$ &Used for comparisons \\
\hline
 Medium (M)                    & 2.3 & $20, 50, 100$ &Used for comparisons \\
\hline
 Dense (D)                     & 2.5 & $20, 50, 100$ &Used for comparisons \\
\hline
 Extreme-dense (XD) & 2.7 &$20, 50, 100$ &Stress test \\
\hline
\end{tabular}
\end{adjustbox}
\end{table}

The instance family identifier ``S'' denotes sparse instances, ``M'' medium-density instances, ``D'' dense instances, and ``XD'' extremely dense stress-test instances. The XD family is included to test solver behaviour under challenging connectivity, not to represent the main operating setting. Regarding the edge factor, it determines the number of edges generated in the graph. A higher edge factor results in a greater number of constraint edges, thereby increasing the density of the benchmark instance and typically making it more difficult to solve.

For each graph size and family, multiple graph seeds generate different structures, and multiple initial-assignment seeds generate different starting assignments for the same planted solution. This separates the effects of graph structure from those of the initial conflict configuration.

Overall, the benchmark comprises 72 instances: 3 graph sizes across 4 families, with 3 graph seeds and 2 initial-assignment seeds per configuration.

Planted satisfying assignments are crucial for interpreting the results: since every instance has at least one valid solution, the success rate reflects the solver’s ability to find a satisfying assignment within the given search budget.

\subsection{Selected algorithms for testing and comparison}
\label{subsec:compared_methods}

We compare 5 algorithms, including the DCC-Incomplete solver under two configurations and 3 state-of-the-art baselines. A list with algorithmic details is provided below.

\begin{itemize}
    \item \textbf{DCC-Incomplete-random}: the proposed 
    solver configured with a random local restart for stagnation recovery.This facilitates assessing the contribution introduced by the ``memory\_guided'' module.
    
    \item \textbf{DCC-Incomplete}: the proposed version using memory-guided restart for stagnation recovery.

    \item \textbf{Distributed Stochastic Algorithm (DSA)} \citep{zhang2005distributedStochastic}: an incomplete 
    distributed local-search algorithm in which agents probabilistically change 
    values according to local improvement.
    
    \item \textbf{Maximum Gain Message algorithm (MGM)} \citep{maheswaran2004mgm}: an incomplete 
    distributed local-search algorithm in which agents exchange ``gain'' information 
    and move only when they have the highest local gain.
    
    \item \textbf{Generalized Distributed Breakout Algorithm (GDBA)} \citep{okamoto2016gdba}: a breakout-based distributed method that adapts constraint weights to escape 
    local minima.
\end{itemize}

These baselines were chosen as established and incomplete distributed constraint-reasoning methods based on different working mechanism. DSA and MGM are classical distributed local-search baselines, while GDBA is a more advanced breakout-style baseline. 

\subsection{Evaluation Metrics}
\label{subsec:evaluation_metrics}

We consider both solution-quality and efficiency metrics. A run is successful if the final cost is zero, indicating a satisfying assignment in the graph-colouring DCSP. If the final cost is greater than zero, the run is unsuccessful under the given stopping limit.

The evaluation metrics are

\begin{itemize}
    \item \textbf{Success rate} (SR): percentage of instances with final null score.
    
    \item \textbf{Runtime} ($\rho$): the wall-clock time (in seconds) required to terminate.
    
    \item \textbf{Iter.}: the number of repairs that occurred before the problem was resolved or the maximum allowed number of iterations \(T_{MAX}\) was reached. 
    
    \item \textbf{Messages} ($N_M$): the total number of explicit communication events recorded during solving.
    
    \item \textbf{Messages per variable} ($N_{M/V}$): the total number of messages normalised by the number of variables.
    
    \item \textbf{Messages per constraint} ($N_{M/C}$): the total number of messages normalised by the number of constraints.

    \item \textbf{Number of Restart} ($N_{R}$): the number of restarts triggered during a run.
    
    \item \textbf{Final cost} (FC): the number of unsatisfied constraints at termination. 

    \item \textbf{Solved cases} ($N_{SC}$): the number of instances solved successfully.

\end{itemize}

For outlier-sensitive metrics (runtime, iterations, messages, messages per variable, and messages per constraint), median values are reported, since unsuccessful or difficult runs can sometimes produce unusually large communication or runtime values (outliers). Final cost (FC) is also reported using the median in the main comparison tables. For the external baselines, DSA, MGM, and GDBA are each executed 10 times over the complete benchmark. Within each execution, success rate is calculated over the corresponding benchmark cases, while the remaining numerical metrics are reported as medians. The final baseline values are then obtained by averaging the 10 execution-level results. When specifically discussing the internal effect of memory-guided restart (Table \ref{tab:summary_comparison_random_memory_restart}, and \ref{tab:stopping_limit_sensitivity}), mean FC (denote as $\overline{FC}$) is additionally reported because it summarises the average remaining conflict level across all runs and highlights the reduction in residual violations. The number of restarts, denoted by $N_R$, and the number of solved cases, denoted by $N_{SC}$, are additionally reported where relevant to the internal DCC-Incomplete analysis. 

\subsection{Experimental settings and considerations}
\label{subsec:experimental_protocol}

All methods are evaluated on the same generated benchmark instances where applicable. For the DCC-Incomplete variants, we use identical graph structures, planted satisfying assignments, and initial assignments, allowing a direct comparison of the restart mechanism by evaluating the random-restart and memory-guided variants under identical conditions.

DSA, MGM, and GDBA are executed using the pyDCOP 
framework\footnote{\url{https://pydcop.readthedocs.io/en/latest/index.html}}. The same graph-colouring instances are converted into the required distributed constraint-reasoning format. A run is successful when the final cost is zero, matching the graph-colouring DCSP setting where zero cost means no inequality constraint is violated.

We run DSA, MGM, and GDBA 10 times per problem to assess reliability (standard deviation).  

The comparison is interpreted cautiously because the proposed solver and the baselines use different internal execution models. DCC-Incomplete relies on repair iterations over a dynamic conflict cluster, whereas the other methods use solver cycles defined by their own algorithms and frameworks. Thus, iteration and cycle counts are reported for transparency but are not the primary criteria for assessing solver quality.

Message counts are interpreted cautiously. DCC-Incomplete records explicit messages from its coordination and repair procedures, while the chosen state-of-the-art methods implement different mechanisms that may alter exact counts. Therefore, communication results are analysed mainly via overall trends and normalised measures such as messages per variable and per constraint.

 DCC-Incomplete, DCC-Incomplete-random, DSA, and MGM were evaluated using a stopping budget of 150 iterations or cycles ($T_{max}=150$), where applicable. GDBA was evaluated using a strict 10-second timeout, because the installed pyDCOP implementation does not support \texttt{stop\_cycle} for GDBA. We remark that the complete source code used to generate these results is made publicly available in \citep{bib:ZenodoDCC-Incomplete}.

\section{Experimental Results}
\label{sec:experimental_results}

\subsection{Functional Validation of DCC-Incomplete}\label{sec:Experimental_Evaluation}

We first test DCC-Incomplete -- overall results are summarised in Table \ref{tab:functional_validation_results} -- on small controlled instances to verify that the solver behaves according to the intended  design and confirm the functional correctness of the main repair and coordination mechanisms.  

\begin{table}[ht!]
\centering
\caption{Summary of functional validation results.}
\label{tab:functional_validation_results}
\begin{adjustbox}{max width=\textwidth}
\begin{tabular}{llll}
\hline
\textbf{Test} & \textbf{Scenario} & \textbf{Mechanism validated} & \textbf{Result} \\
\hline
Test 1 & Three-variable conflict & Single conflict cluster formation and accepted repair & Passed \\
\hline
Test 2 & Two separated conflict chains & Independent conflict cluster formation & Passed \\
\hline
Test 3 & Boundary-conflicting proposal & Deferred proposal and postponed merge request & Passed \\
\hline
Test 4A & Stagnation with sideways move & Sideways stagnation recovery & Passed \\
\hline
Test 4B & Stagnation without sideways move & Local restart recovery & Passed \\
\hline
Test 5 & Postponed merge-request mechanism & Postponed queue handling and rechecked merge request & Passed \\
\hline
\end{tabular}
\end{adjustbox}
\end{table}

The algorithms passed the evaluation (i.e., successfully solved the simple problems), allowing us to confirm that the core mechanisms in DCC-Incomplete behaved correctly. The solver is able to form conflict clusters, coordinate local repairs, defer proposals that conflict at boundaries, postpone merge requests, reconstruct conflict structures, and recover from stagnation. Detailed information on each individual test is provided below.

\subsubsection{Detailed analysis of Test 1}
\label{subsec:preliminary_validation}

Test 1 is a small graph-colouring DCSP with three agents, three variables, and two binary inequality constraints, $c_{12}: x_1 \neq x_2$ and $c_{23}: x_2 \neq x_3$. With a fixed random seed of 15, the initial assignment was $\alpha_0 = \{x_1 = R,\; x_2 = R,\; x_3 = R\}$. Both constraints were violated, yielding scores $\mathrm{Score}(a_1) = -1$, $\mathrm{Score}(a_2) = -2$, and $\mathrm{Score}(a_3) = -1$.

All three agents were classified as active and formed one conflict cluster. Agent $a_1$ became the leader by the least-conflicted-agent rule, with the smallest identifier breaking ties. Agent $a_2$, having the lowest score, was chosen as the target. It proposed changing $x_2$ from $R$ to $G$, reducing its local cost from 2 to 0. As this introduced no boundary conflict, the change was accepted, yielding $\alpha_1 = \{x_1 = R,\; x_2 = G,\; x_3 = R\}$. All constraints were then satisfied, and the reconstructed conflict subgraph became empty.

\begin{figure}[!]
    \centering
    \includegraphics[width=0.9\textwidth]{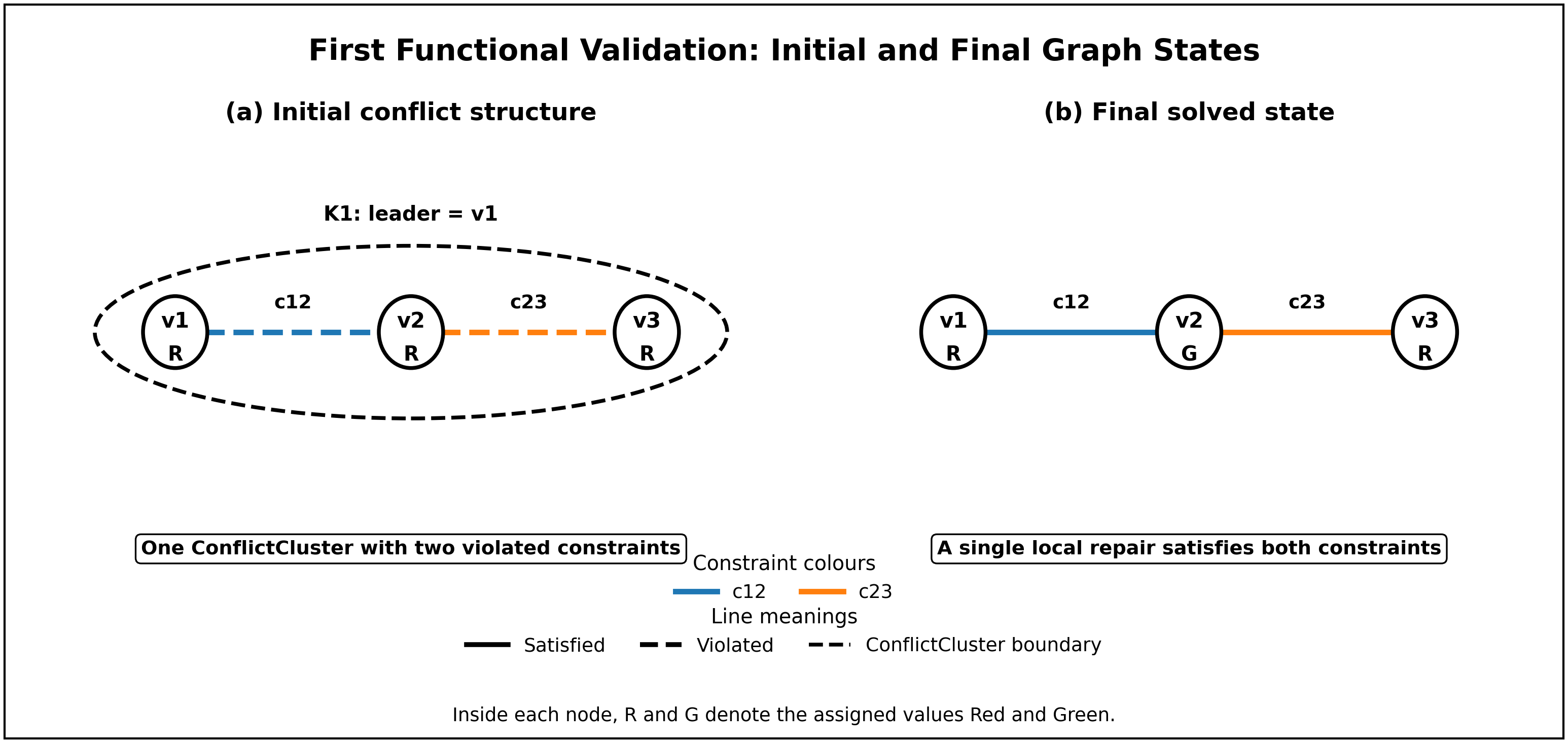}
    \caption{Initial and final graph states for the first functional validation}
    \label{fig:test1_initial_and_final_graph_states}
\end{figure}

Figure \ref{fig:test1_initial_and_final_graph_states} shows the initial and final graph states of the first controlled validation instance. Initially, all three variables are set to \(R\), yielding two violated constraints and one conflict cluster (\(K_1=\{v_1,v_2,v_3\}\)).

\begin{figure}[ht]
    \centering
    \includegraphics[width=0.9\textwidth]{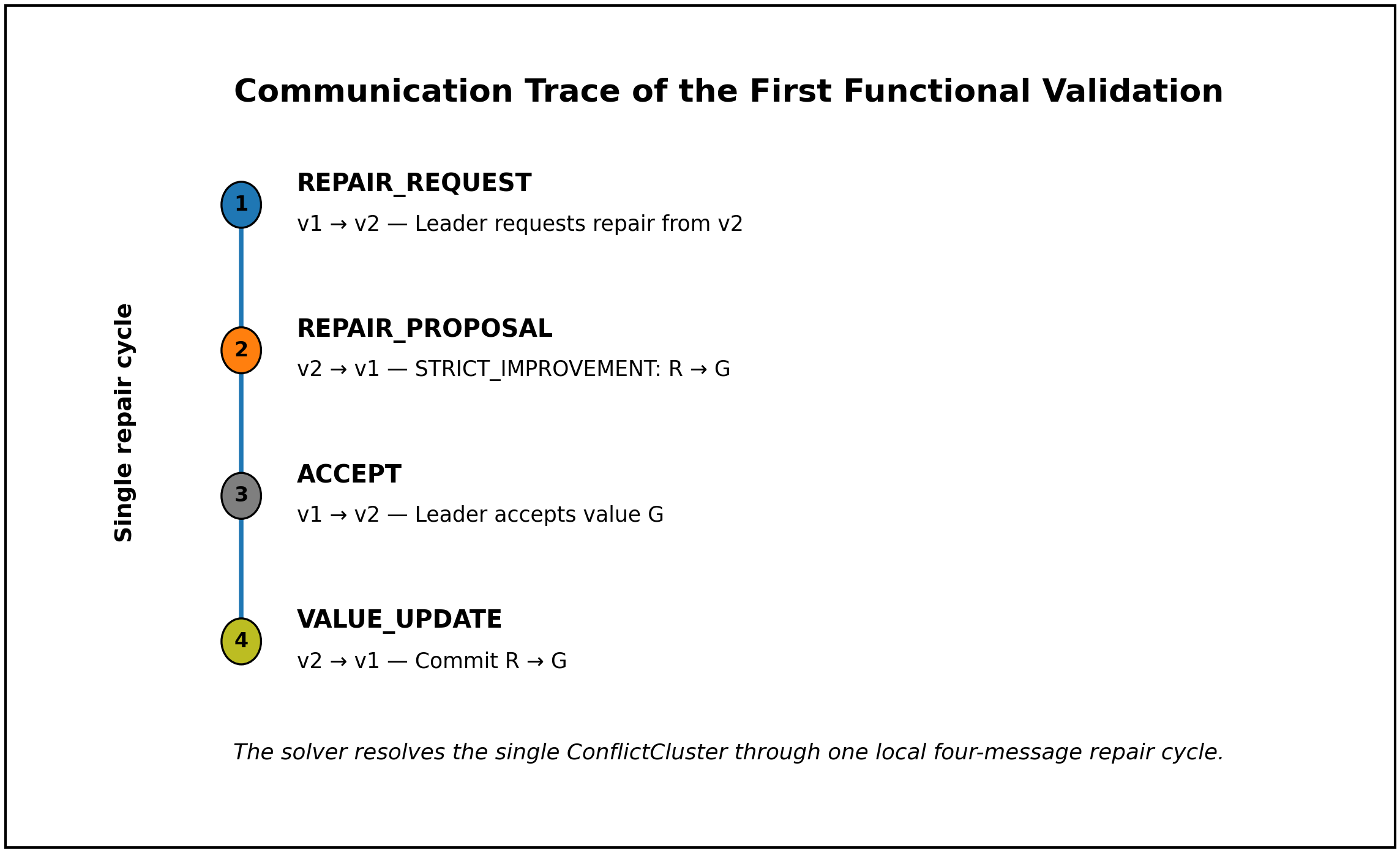}
    \caption{Communication trace for the Test 1.}
    \label{fig:test1_communication_trace}
\end{figure}

The leader (\(v_1\)) requests a repair proposal from \(v_2\). The accepted update (\(v_2:R\rightarrow G\)) resolves both violations, yielding the satisfying assignment \(\{v_1=R,\;v_2=G,\;v_3=R\}\). Figure \ref{fig:test1_communication_trace} shows the communication trace. The repair uses one local four-message exchange: \texttt{REPAIR\_REQUEST}, \texttt{REPAIR\_PROPOSAL}, \texttt{ACCEPT}, and \texttt{VALUE\_UPDATE}.

The repair required four explicit messages: `REPAIR\_REQUEST', `REPAIR\_PROPOSAL', `ACCEPT', and `VALUE\_UPDATE'. This confirms that the algorithm performs conflict detection, dynamic classification, cluster construction, coordinated repair, 
assignment update, and local reconstruction correctly.

Table \ref{tab:preliminary_validation} summarises the preliminary validation results from the prototype output, including the initial conflict state, selected leader and target agents, accepted repair action, communication messages, and final assignment.

\begin{table}[!] \centering \caption{Test 1 in numbers.} \label{tab:preliminary_validation} 
\small
\begin{tabular}{ll}
\hline
\textbf{Metric} & \textbf{Result} \\ 
Initial violated constraints & 2 \\ 
Initial  \texttt{Active Agents} & 3 \\ 
Initial \texttt{conflict clusters} & 1 \\  
Selected leader & $a_1$ \\ 
Selected target & $a_2$ \\ 
Accepted update & $x_2: R \rightarrow G$ \\  
Examined targets before acceptance & 1 \\  
Explicit communication messages & 4 \\ 
Final violated constraints & 0 \\ 
Final \texttt{Active Agents} & 0 \\  
Final 
\texttt{conflict cluster} & 0 \\ 
\hline Final status & Solved \\ 
\hline 
\end{tabular} \end{table}

\subsubsection{Detailed analysis of Test 2}
\label{subsec:multiple_conflict_clusters}

Test 2 evaluates whether \textit{DCC-Incomplete} correctly identifies and represents multiple independent conflicted regions within a single DCSP instance. We used a six-node graph-colouring problem with six agents and two disconnected chains, $x_1 - x_2 - x_3$ and $x_4 - x_5 - x_6$. Each variable has a binary domain $\{R,G\}$, and adjacent variables must take different values. With a fixed random seed of 15, the initial assignment was $\alpha_0 = \{x_1 = R,\; x_2 = R,\; x_3 = R,\; x_4 = R,\; x_5 = R,\; x_6 = R\}$, which violated four constraints: $c_{12},\; c_{23},\; c_{45},\; c_{56}$. The resulting conflict subgraph contained two independent connected components, 
which were correctly extracted as the two \texttt{conflict cluster}$K_1 = \{a_1, a_2, a_3\}$ and $K_2 = \{a_4, a_5, a_6\}$.

For $K_1$, agent $a_1$ was selected as leader, and for $K_2$, agent $a_4$ was selected. In both clusters, the leader was the least-conflicted member under the proposed rule. Each cluster had a Cluster Conflict Load of 4, representing the agents’ total conflict involvement rather than the number of distinct violated constraints. The initial cluster-level output appears in Table \ref{tab:multi_cluster_initial_state}.

\begin{table}[ht!]
\centering
\caption{Initial state of the multi-cluster validation instance}
\label{tab:multi_cluster_initial_state}
\small
\begin{tabular}{lll}
\hline
\textbf{Conflict cluster} & \textbf{$K_1$} & \textbf{$K_2$} \\
\hline
Cluster members &
$\{a_1, a_2, a_3\}$ &
$\{a_4, a_5, a_6\}$ \\
\hline
Violated constraints &
$\{c_{12}, c_{23}\}$ &
$\{c_{45}, c_{56}\}$ \\
\hline
Selected leader &
$a_1$ &
$a_4$ \\
\hline
Most-conflicted initial target &
$a_2$ &
$a_5$ \\
\hline
Cluster Conflict Load &
$4$ &
$4$ \\
\hline
\end{tabular}
\end{table}

This validation confirms that the solver can dynamically identify multiple \texttt{conflict clusters}, assign each a separate leader, and maintain independent coordination records. The next experiment tests whether the solver repairs these clusters sequentially, reconstructs the dynamic state after each accepted update, and removes solved clusters from the active list. The full solver loop was then run to verify this behaviour, with results shown in Table \ref{tab:multi_cluster_repair_process}.

\begin{table}[ht] \centering \caption{Summary of the complete multi-cluster repair process} \label{tab:multi_cluster_repair_process} 
\begin{adjustbox}{max width=0.9\textwidth}
\begin{tabular}{lll}
\hline 
\textbf{} & \textbf{Initial state} & \textbf{Final state} \\ 
\hline
Number of violated constraints & $4$ & $0$ \\ 
Number of \texttt{Active Agents} & $6$ & $0$ \\ 
Number of \texttt{conflict cluster} & $2$ & $0$ \\ 
Accepted value updates & --- & $x_2: R \rightarrow G;\quad x_5: R \rightarrow G$ \\ 
Number of solver iterations & --- & $2$ \\ 
Total explicit messages & --- & $8$ \\ 
\hline
Final solver status & --- & Solved \\ 
\hline \end{tabular}
\end{adjustbox}
\end{table}

The results confirm that the solver correctly applies localised repair, reconstructs the dynamic conflict structure after each accepted update, and terminates when no constraints are violated.

\begin{figure}[ht]
    \centering
    \includegraphics[width=0.9\textwidth]{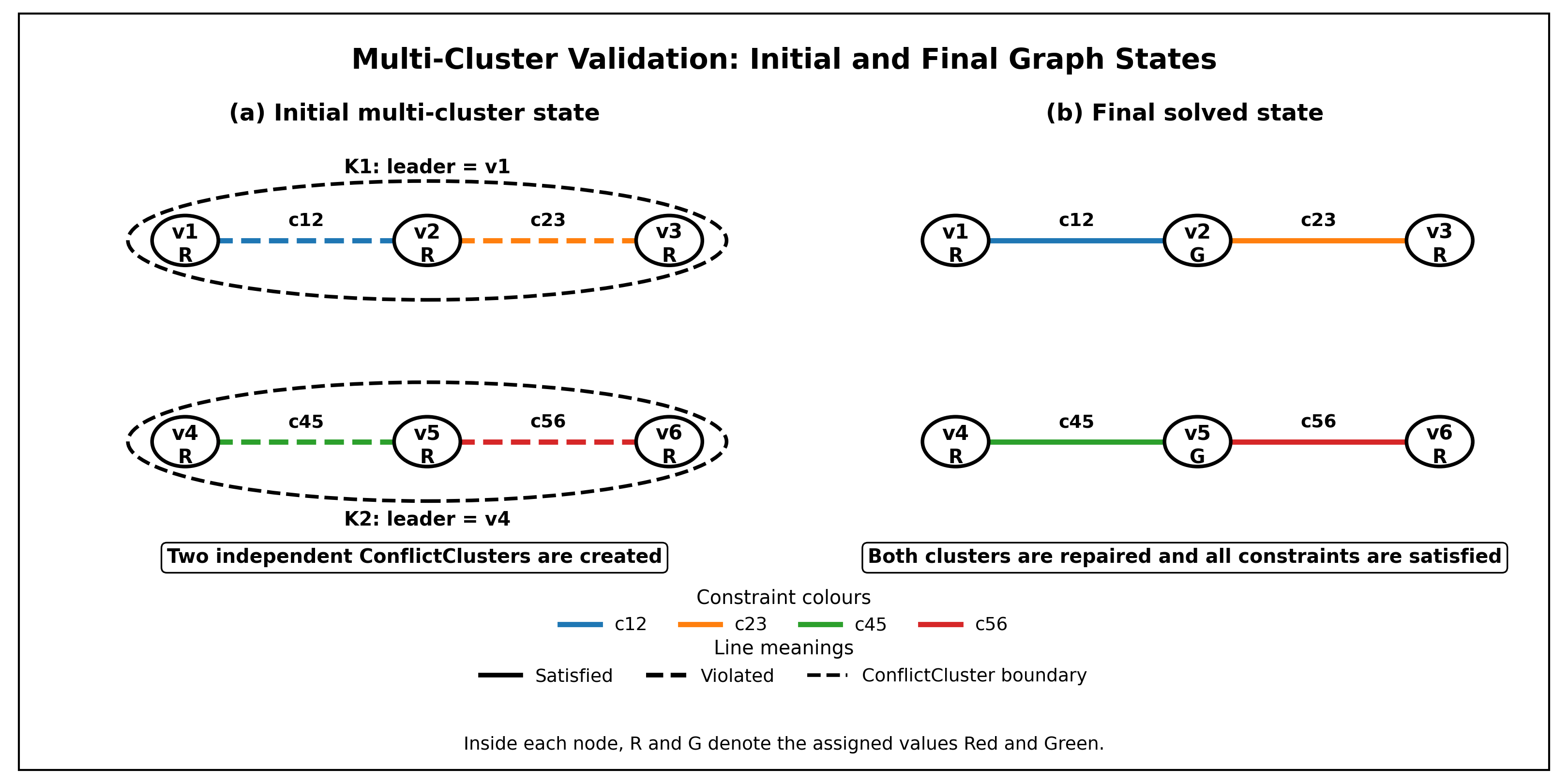}
    \caption{Multi-Cluster Validation: Initial and Final Graph States}
    \label{fig:problem2_initial_and_final_states}
\end{figure}

Figure \ref{fig:problem2_initial_and_final_states} shows the multi-cluster validation instance before and after repair. Initially, two independent conflict clusters each contain a local violation; in the final state, both clusters are successfully repaired and all constraints are satisfied.

\begin{figure}[ht!]
    \centering
    \includegraphics[width=0.8\textwidth]{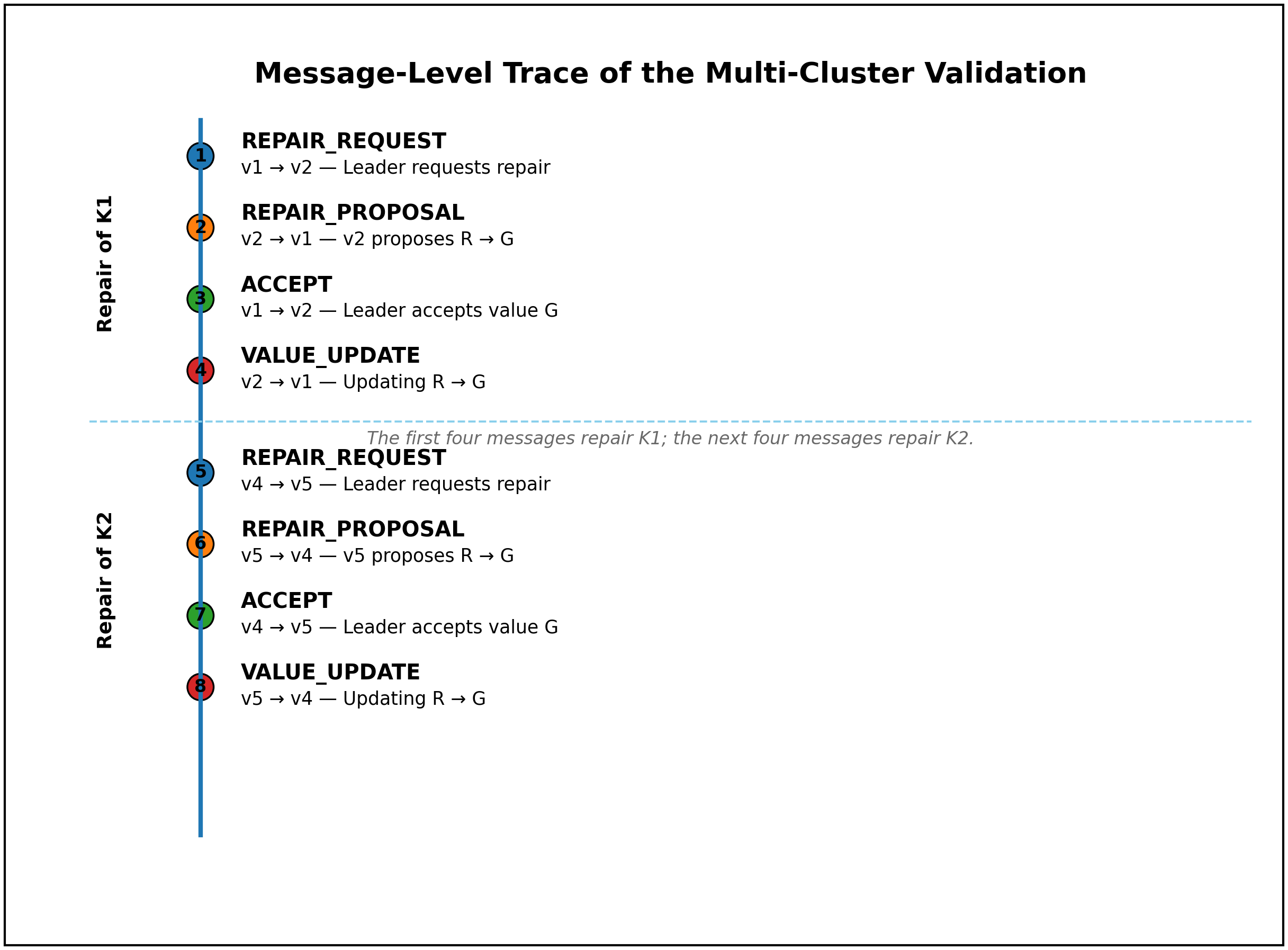}
    \caption{The Communication Trace of the Multi-cluster Validation Test}
    \label{fig:problem2_message_trace}
\end{figure}

Figure \ref{fig:problem2_message_trace} shows the communication trace of the multi-cluster validation test: the solver repairs each conflict cluster with a local four-message exchange, confirming localised repair and structured communication.

\subsubsection{Detailed analysis of Test 3}
\label{subsec:boundary_conflict_validation}

Test 3 verifies how a locally improving proposal is handled when it creates a boundary conflict with an adjacent \texttt{conflict cluster}. We used a controlled five-agent graph-colouring instance with four binary inequality constraints, \(c_{12}:x_1\neq x_2\), \(c_{23}:x_2\neq x_3\), \(c_{24}:x_2\neq x_4\), and \(c_{45}:x_4\neq x_5\). Equivalently, the graph
contains two connected paths, \(x_1-x_2-x_3\) and \(x_2-x_4-x_5\), which share
the variable \(x_2\). The domain is the same as in Test~2. With a fixed random
seed of 15, the initial assignment was \(\alpha_0=\{x_1=R,x_2=R,x_3=R,x_4=G,x_5=G\}\), with violated constraints \(c_{12}\), \(c_{23}\), and \(c_{45}\), and two resulting conflict clusters:
\(K_1=\{a_1,a_2,a_3\}\) and \(K_2=\{a_4,a_5\}\). The initial cluster configuration is summarised in Table~\ref{tab:boundary_conflict_initial_state}.

\begin{table}[ht!]
\centering
\caption{Initial state of the boundary-conflict validation instance}
\label{tab:boundary_conflict_initial_state}
\renewcommand{\arraystretch}{1.2}
\small
\begin{tabular}{lll}
\hline
\textbf{Conflict cluster} & \textbf{$K_1$} & \textbf{$K_2$} \\
\hline
Cluster members &
$\{a_1, a_2, a_3\}$ &
$\{a_4, a_5\}$ \\
Initially violated constraints &
$c_{12}, c_{23}$ &
$c_{45}$ \\
Selected leader &
$a_1$ &
$a_4$ \\
First repair target &
$a_2$ &
$a_5$ \\
Cluster conflict load &
$4$ &
$2$ \\
Boundary edge &
$(a_2, a_4)$ through $c_{24}$ &
$(a_4, a_2)$ through $c_{24}$ \\
\hline
\end{tabular}
\end{table}

In the first repair cycle, agent $a_2$ proposed changing $x_2$ from $R$ to $G$, lowering its local cost from 2 to 1. Although locally improving, this created a boundary conflict with $a_4$ via constraint $c_{24}$, so the proposal was deferred. The leader of $K_1$ then requested a merge with the leader of $K_2$. After acceptance, the deferred update was committed, the deferred queue cleared, and the dynamic state reconstructed. The new structure selected $a_2$ as leader, and a second repair changed $x_4$ from $G$ to $R$. The final satisfying assignment was: $\alpha_f =
    \{x_1 = R,\; x_2 = G,\; x_3 = R,\; x_4 = R,\; x_5 = G\}$.

The overall outcome is summarised in 
Table~\ref{tab:boundary_conflict_summary}. The results confirm that the prototype correctly detects a boundary conflict, 
stores the corresponding proposal in the deferred queue, activates 
leader-to-leader coordination, commits the deferred update after merge 
acceptance, clears outdated deferred proposals, reconstructs the dynamic state, 
reselects the leader, and terminates after the remaining conflict is repaired.

\begin{table}[ht!]
\centering
\caption{Summary of the boundary-conflict and merge-handling validation phase.}
\label{tab:boundary_conflict_summary}
\small
\begin{tabular}{ll}
\hline
\textbf{Metric} & \textbf{Result} \\
\hline
Initial violated constraints &
$3$ \\
Initial \texttt{conflict cluster} &
$2$ \\
Deferred proposals &
$1$ \\
Leader-to-leader messages &
$2$ \\
Accepted value updates &
$x_2: R \rightarrow G;\quad x_4: G \rightarrow R$ \\
Solver iterations &
$2$ \\
Total explicit messages &
$13$ \\
Final violated constraints &
$0$ \\
Final solver status &
Solved \\
\hline
\end{tabular}
\end{table}

\begin{figure}[ht!]
    \centering
    \includegraphics[width=0.9\textwidth]{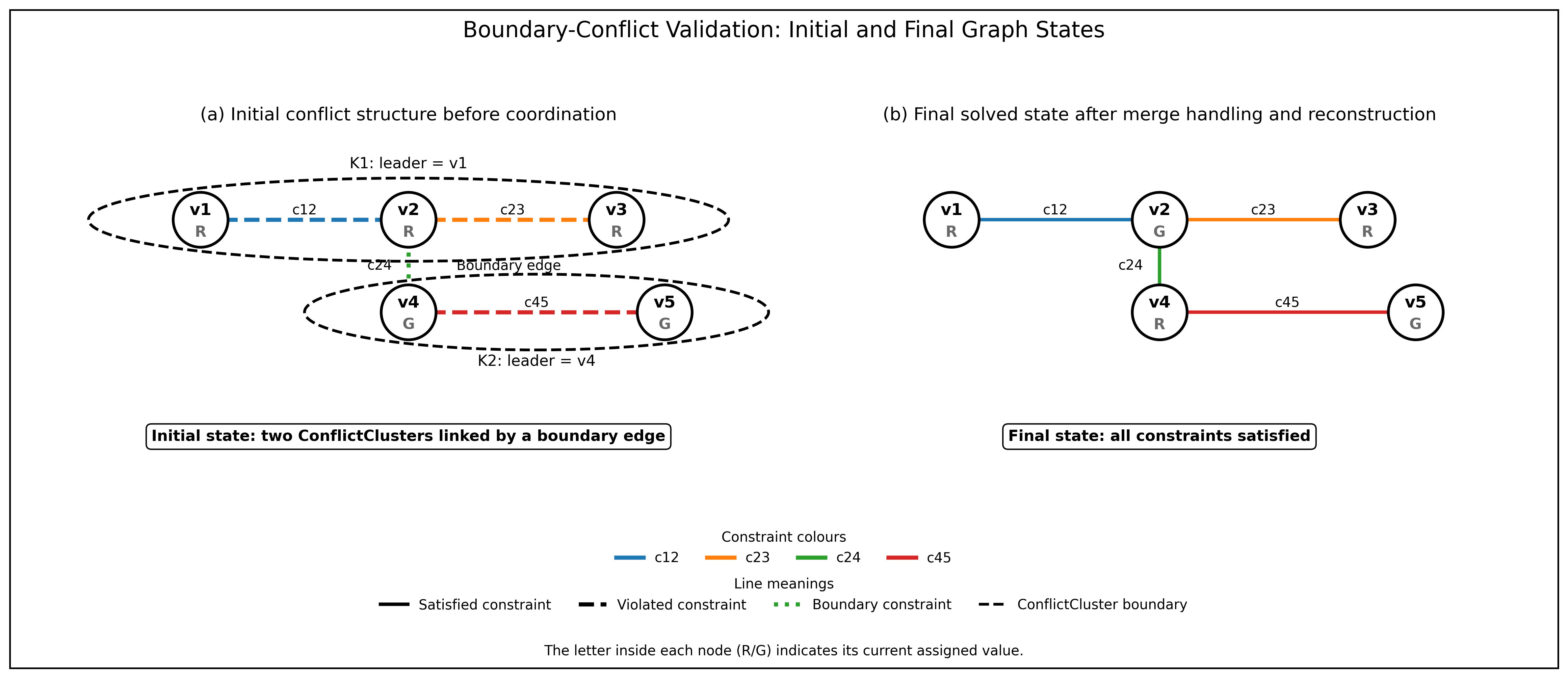}
    \caption{Test 3 initial conflict structure of the controlled boundary-conflict validation instance (a) and final state after merge handling and dynamic reconstruction (b).}
    \label{fig:stage3_initial_and_final_states}
\end{figure}

In figure \ref{fig:stage3_initial_and_final_states} the sate (a) shows the initial conflict structure of the controlled boundary-conflict validation instance. The violated constraints produce two \texttt{conflict cluster}, $K_1 = \{a_1, a_2, a_3\}$ and 
$K_2 = \{a_4, a_5\}$, connected by the boundary constraint $c_{24}$. State (b)  visualise the final state after merge handling and dynamic reconstruction. 
The accepted updates $x_2: R \rightarrow G$ and $x_4: G \rightarrow R$ eliminate all constraint violations.

\begin{figure}[ht]
    \centering
    \includegraphics[width=0.9\textwidth]{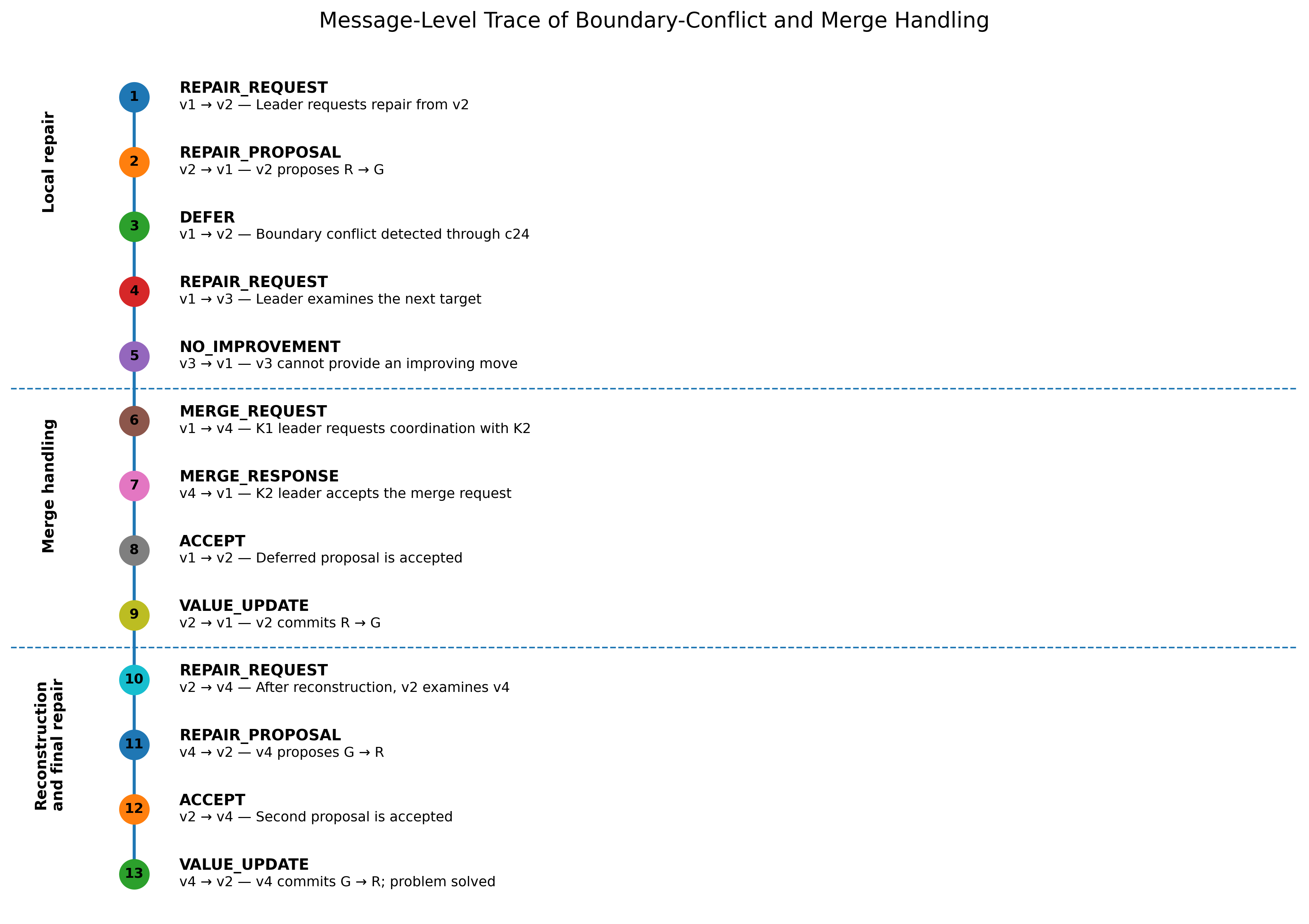}
    \caption{Message-level trace of the boundary-conflict validation}
    \label{fig:stage3_coordination_timeline}
\end{figure}

Figure \ref{fig:stage3_coordination_timeline} shows the message-level trace of the boundary-conflict validation, including the deferred proposal, leader-to-leader merge coordination, accepted update, dynamic reconstruction, and final repair action.

A locally improving proposal is deferred because it introduces a boundary conflict via $c_{24}$. After leader-to-leader coordination and merge acceptance, the update is committed, the dynamic state rebuilt, and a repair then yields a satisfying assignment.

\subsubsection{Detailed analysis of Test 4A and 4B}

With two tests, we verify stagnation handling: the first (Test 4A) evaluates recovery via an equal-cost sideways move; the second (Test 4B) evaluates recovery via a local restart when no strictly improving or acceptable sideways move exists.

\begin{figure}[ht]
    \centering
    \includegraphics[width=0.8\textwidth]{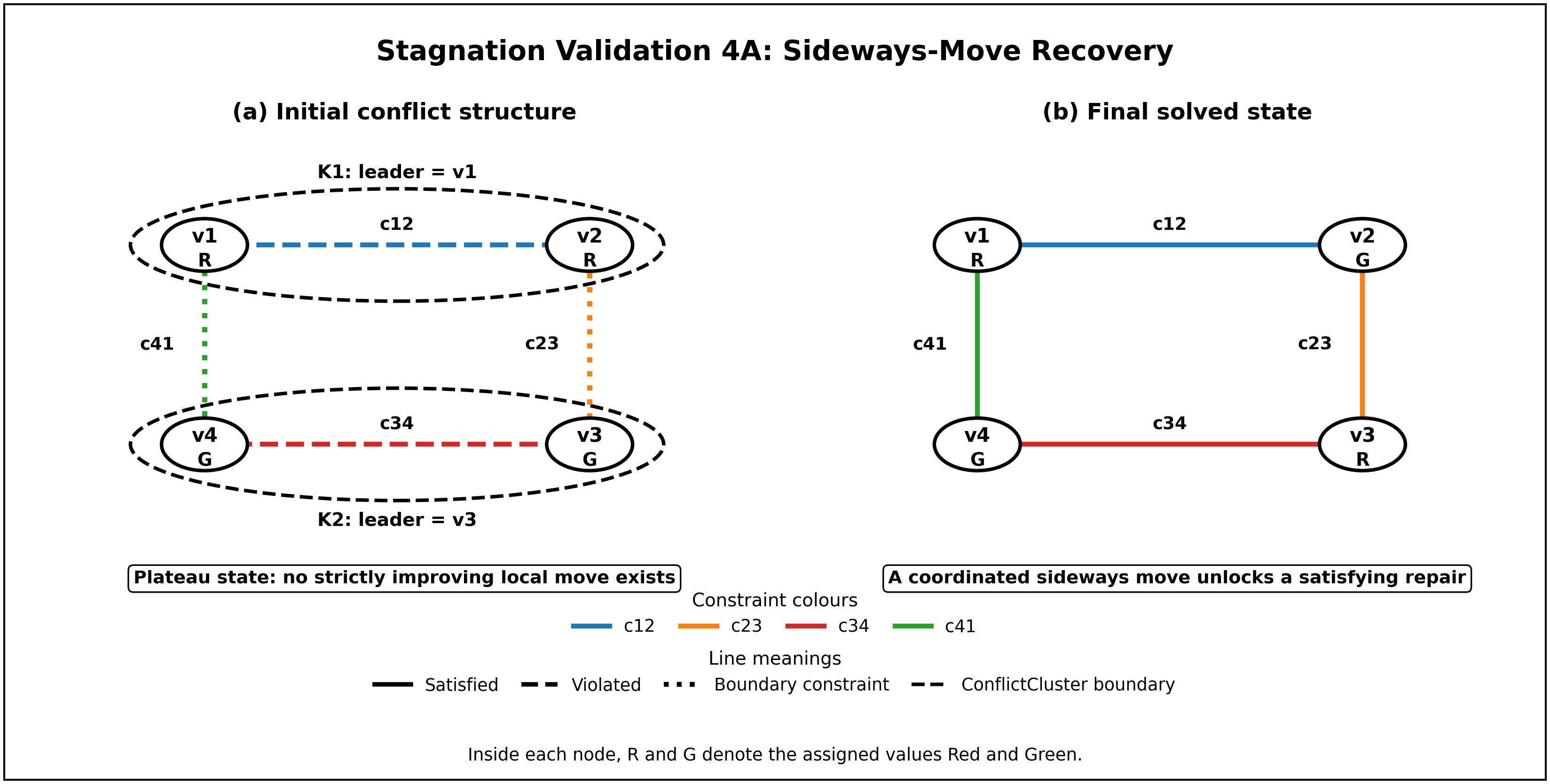}
    \caption{Initial and final graph states for sideways-move recovery (Test 4A).}
\label{fig:stage4_testA_sideways_initial_and_final_states}
\end{figure}

For Test~4A, with fixed solver seed 15 and the initial assignment \(\alpha_0=\{v_1=R,v_2=R,v_3=G,v_4=G\}\), the initially violated constraints were \(c_{12}\) and \(c_{34}\). The corresponding initial conflict clusters were \(K_1=\{v_1,v_2\}\) and \(K_2=\{v_3,v_4\}\), and
\(\tau\) was set to 2. Each cluster first plateaus because no strictly improving move exists. After repeated failures of the local-repair procedure for $K_1$, its \texttt{NoImproveCount} reaches 2, triggering stagnation mode and allowing equal-cost sideways moves. Agent $v_2$ proposes the update $v_2:R\rightarrow G$, which preserves its local cost ($1\rightarrow1$). However, this move causes a boundary conflict with $v_3$ via $c_{23}$, so the proposal is deferred. The leader of $K_1$ coordinates with the leader of $K_2$, and the update is committed only after the merge request is accepted. After dynamic reconstruction, a strictly improving update becomes available and the solver finds a satisfying assignment. Figure \ref{fig:stage4_testA_sideways_initial_and_final_states} represents the initial and final graph states of Test 4A. The initial state has two conflict clusters and no strictly improving repair. After stagnation detection, the coordinated sideways update $v_2:R\rightarrow G$ is applied, and a subsequent strict repair yields a satisfying assignment.

\begin{figure}[ht]
    \centering
    \includegraphics[width=0.65\textwidth]{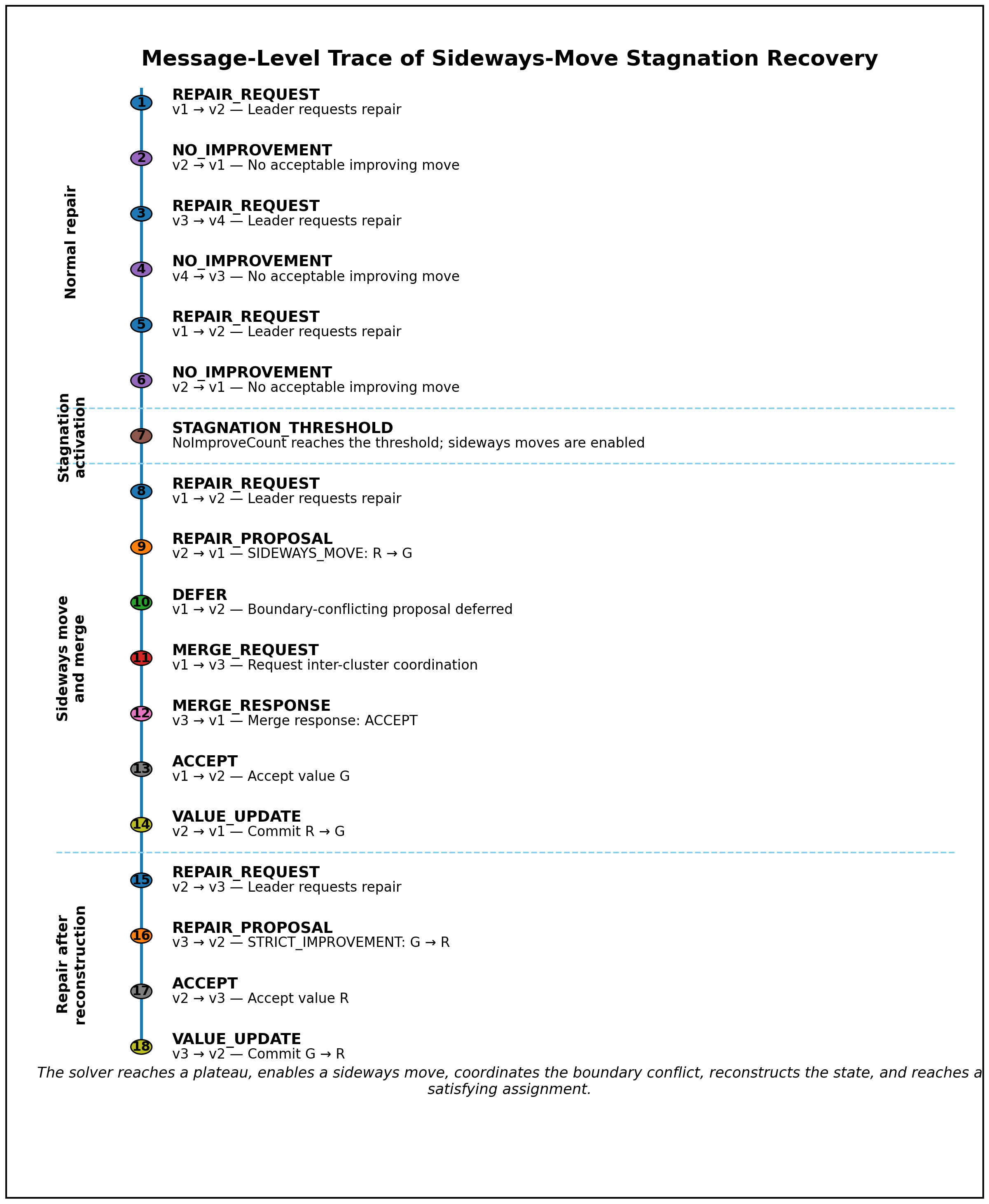}
    \caption{Communication trace for sideways-move recovery (Test 4A).}
    \label{fig:stage4_testA_sideways_message_traces}
\end{figure}

Figure \ref{fig:stage4_testA_sideways_message_traces} shows the event-level trace, including both explicit messages and the internal threshold event that marks when stagnation handling becomes active. The trace shows the normal repair attempts, stagnation-threshold activation, sideways proposal, deferred update, leader-to-leader merge, dynamic reconstruction, and final strict repair.

\begin{figure}[ht!]
    \centering
    \includegraphics[width=0.8\textwidth]{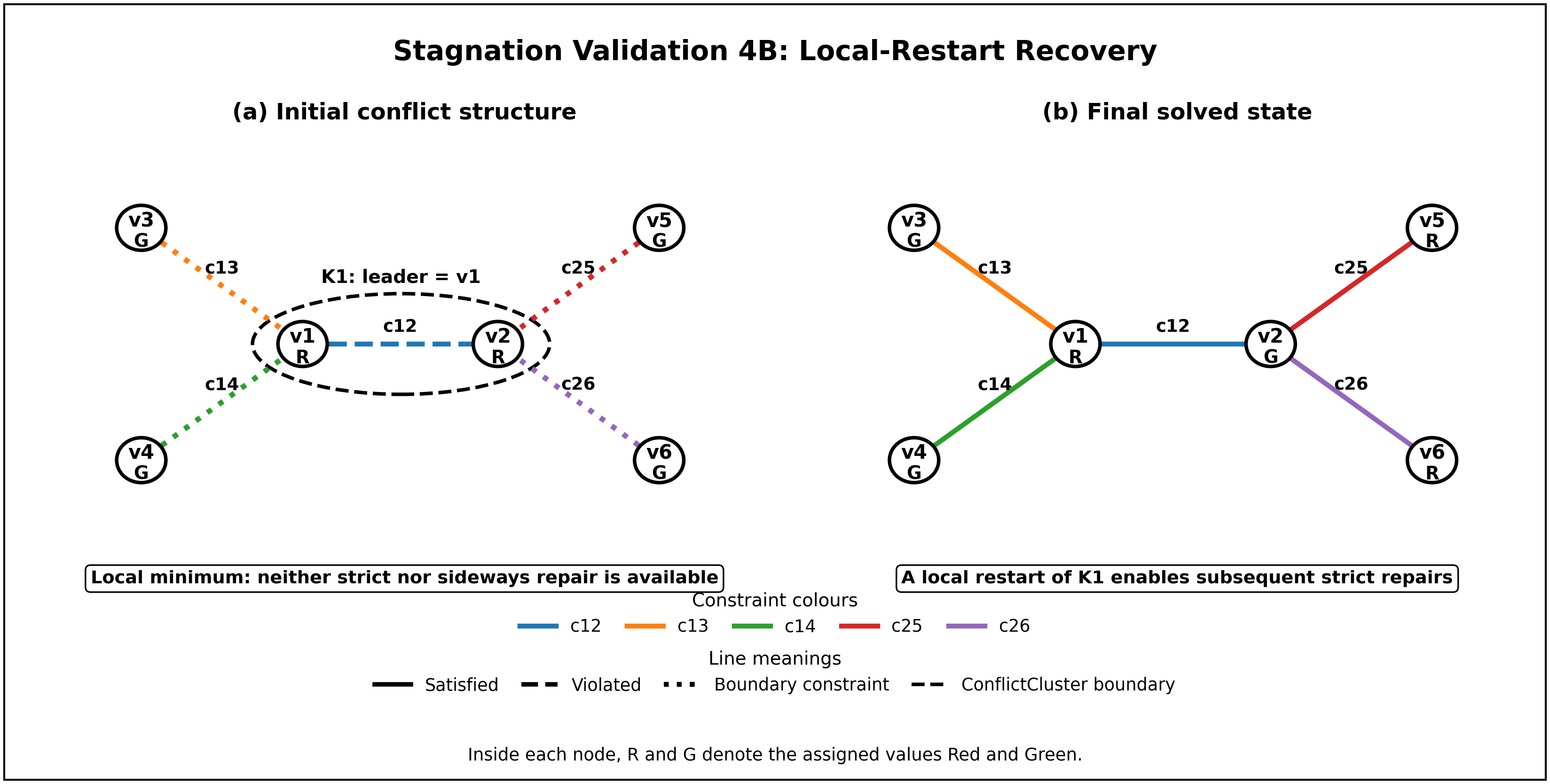}
    \caption{Initial and final graph states (Test 4B).}
    \label{fig:stage4_testB_restart_initial_and_final_states}
\end{figure}

For Test 4B, with fixed solver seed 15 and the initial assignment $\alpha_0 = \{v_1 = R,\; v_2 = R,\; v_3 = G,\; v_4 = G,\; v_5 = G,\; v_6 = G\}$, the only initially violated constraint was $c_{12}$. The corresponding initial conflict cluster was $K_1 = \{v_1, v_2\}$, and $\tau$ was set to $1$. Here, changing $v_1$ or $v_2$ alone would create more conflicts, so the normal repair procedure cannot find a strictly improving move. In stagnation mode, no acceptable equal-cost sideways move exists either, so the solver performs a local restart on the affected conflict cluster. The restart uses seed 16 and changes $\{v_1=R,\;v_2=R\}\rightarrow\{v_1=G,\;v_2=G\}$. Variables outside $K_1$ remain unchanged. The deferred queue is cleared, \texttt{NoImproveCount} is reset to zero, and the dynamic state is rebuilt before repair resumes. Figure \ref{fig:stage4_testB_restart_initial_and_final_states} shows the initial and final graph states. The initial assignment has a local minimum in $K_1=\{v_1,v_2\}$. A local restart reassigns only $K_1$, and the resulting strict repairs yield a satisfying assignment.

\begin{figure}[ht!]
    \centering
    \includegraphics[width=0.8\textwidth]{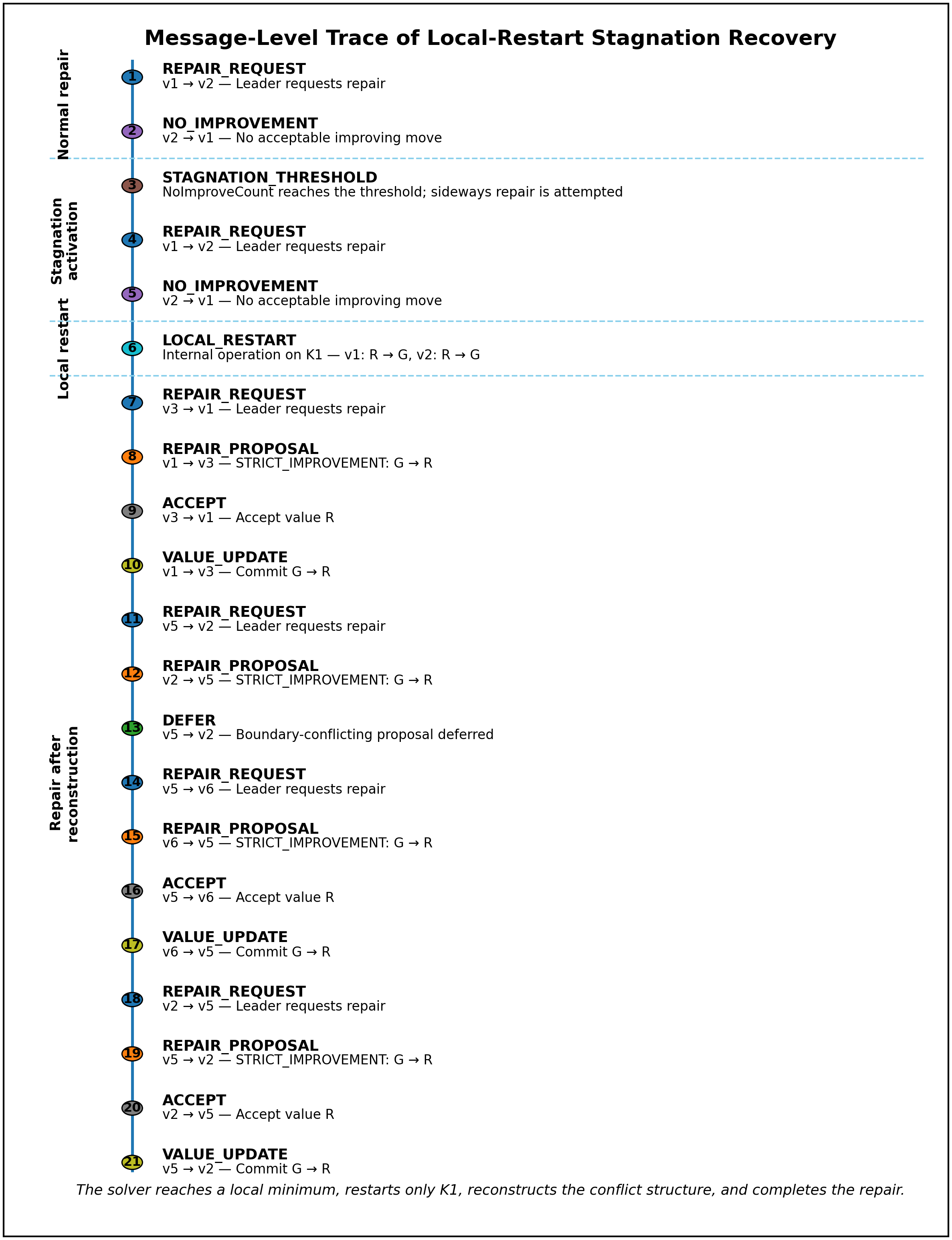}
    \caption{Communication trace (Test 4B)}
    \label{fig:stage4_testB_restart_message_trace}
\end{figure}

Figure \ref{fig:stage4_testB_restart_message_trace} shows the event-level trace. The local restart appears as an internal event even though it does not increase the explicit message count. The trace shows failed normal and sideways repairs, activation of the local-restart branch, dynamic reconstruction, and the strict repairs that complete the solution.

Table \ref{tab:Summary_of_the_stagnation-handling_test} compares the confirmed outcomes of the two stagnation tests.

\begin{table}[ht!]
\centering
\caption{Summary of the stagnation-handling validation results} \label{tab:Summary_of_the_stagnation-handling_test}
\begin{adjustbox}{max width = \textwidth}
\small
\begin{tabular}{lll}
\hline
\textbf{Test} &
\textbf{4A} &
\textbf{4B} \\
\hline
Initial violated constraints & 2 & 1 \\
Stagnation threshold & 2 & 1 \\
Threshold reached at iteration & 2 & 1 \\
Recovery mechanism & Coordinated sideways move & Local restart \\
Sideways update & $v_2:R\rightarrow G$ & Not available \\
Merge coordination required & Yes & No \\
Restarted conflict cluster & --- & $K_1=\{v_1,v_2\}$ \\
Restart seed & --- & 16 \\
Variables outside restarted cluster preserved & --- & Yes \\
Dynamic reconstruction applied & Yes & Yes \\
Local restarts & 0 & 1 \\
Solver iterations & 3 & 4 \\
Explicit messages & 17 & 19 \\
Final violated constraints & 0 & 0 \\
Final solver status & Solved & Solved \\
\hline
\end{tabular}
\end{adjustbox}
\end{table}

The results show that the solver escapes local stagnation in two ways. If an equal-cost sideways move exists, it is still checked for boundary conflicts and coordinated between leaders before commitment. If no improving or sideways move is available, the solver restarts only the affected conflict cluster, clears stale information, rebuilds the dynamic state, and resumes local repair until it finds a satisfying assignment.

\subsubsection{Detailed analysis of Test 5}

A final test evaluates the postponed merge-request mechanism. This mechanism is used when a boundary-conflicting proposal cannot be processed because the neighbouring conflict cluster is busy. The merge request is then postponed, stored, and retried once the neighbouring cluster is available.

\begin{figure}[ht]
    \centering
    \includegraphics[width=0.8\textwidth]{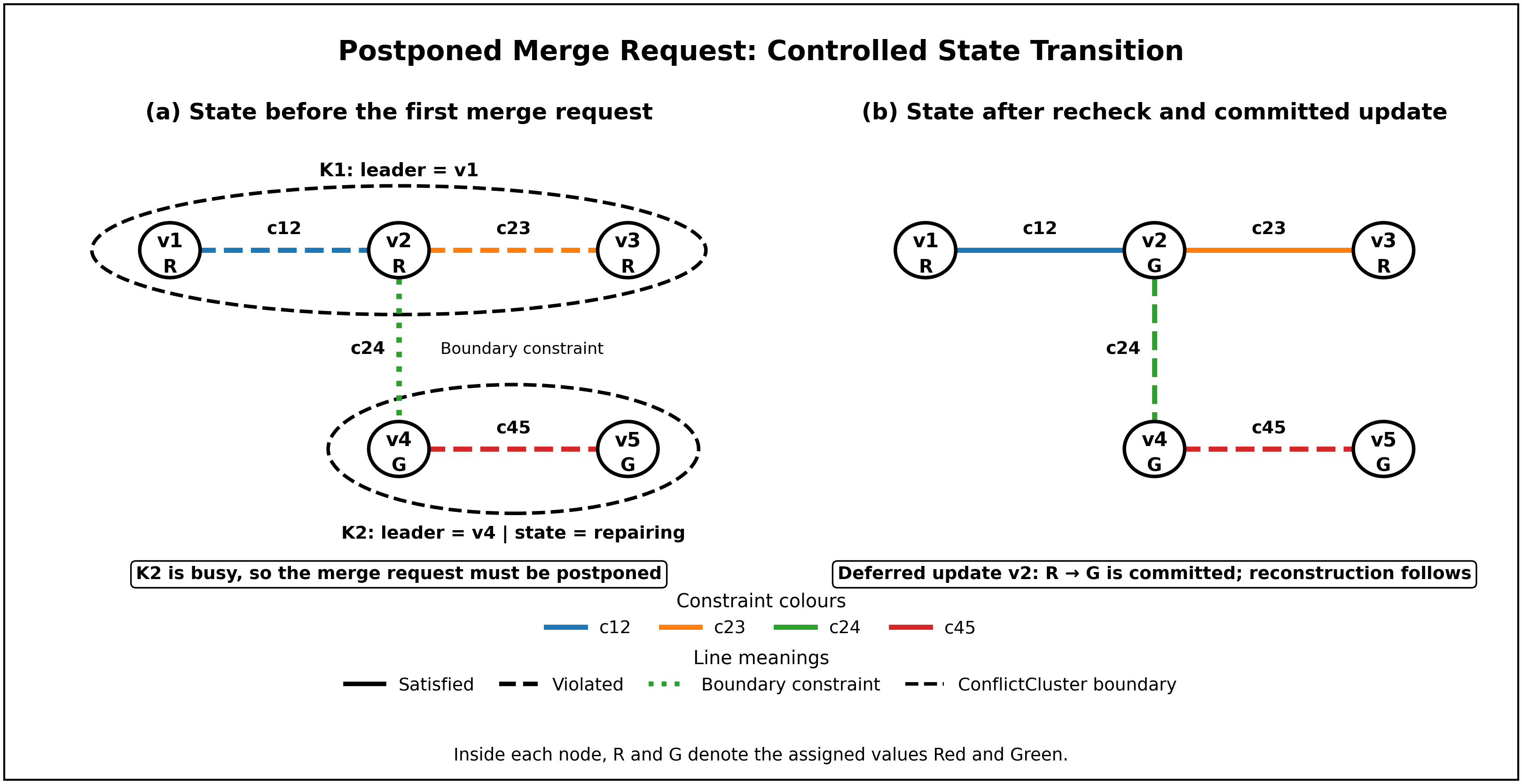}
    \caption{Initial and updated graph states (Test 5).}
    \label{fig:stage5_postponed_merge_initial_and_updated_states}
\end{figure}

The controlled graph-colouring instance has five variables and four binary inequality constraints. We used the fixed initial assignment \(\alpha_0=\{v_1=R,v_2=R,v_3=R,v_4=G,v_5=G\}\) and solver seed 15 for Test 5. The reconstructed conflict structure had $K_1=\{v_1,v_2,v_3\}$ and $K_2=\{v_4,v_5\}$, with leaders \(v_1\) and \(v_4\), respectively. The proposal $v_2:R\rightarrow G$ was linked to boundary constraint \(c_{24}\), which connects \(v_2\) and \(v_4\). To activate the postponed-request branch, \(K_2\) was temporarily set to the \texttt{repairing} state. Figure \ref{fig:stage5_postponed_merge_initial_and_updated_states} shows the state before the first merge request and after the postponed request is rechecked and accepted. When the initial \texttt{MERGE\_REQUEST} arrives, \(K_2\) is busy and replies \texttt{MERGE\_RESPONSE: POSTPONE}. The assignment stays the same and the request is queued. Once \(K_2\) is idle, the queued request is retried, the neighbouring leader replies \texttt{MERGE\_RESPONSE: ACCEPT}, and the deferred update \(v_2:R\rightarrow G\) is committed.

\begin{figure}[ht]
    \centering
    \includegraphics[width=0.8\textwidth]{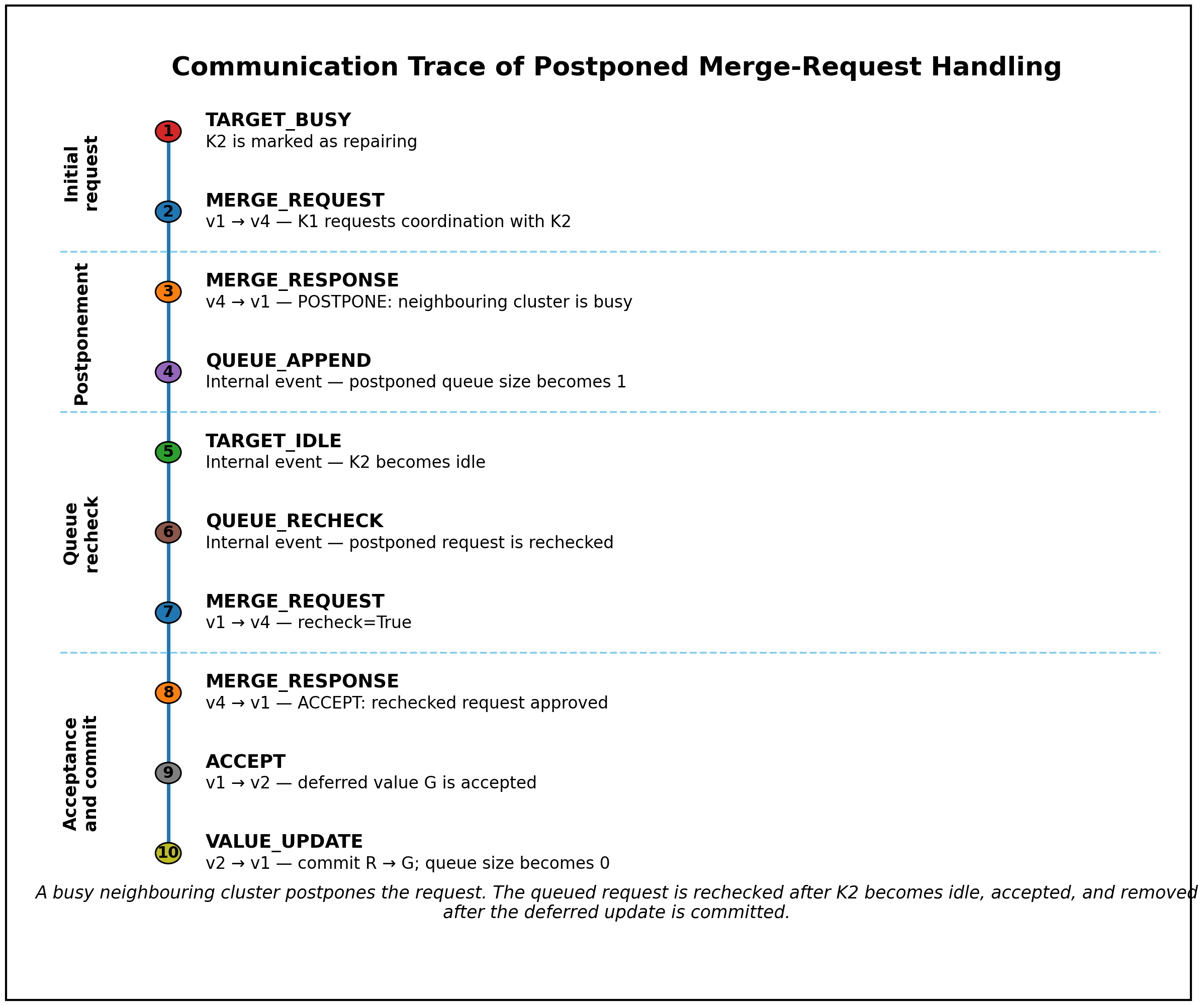}
    \caption{Communication trace (Test 5).}
    \label{fig:stage5_postponed_merge_communication_trace}
\end{figure}

Figure \ref{fig:stage5_postponed_merge_communication_trace} shows the obtained communication trace. The confirmed updated assignment is: $ \alpha'= \{v_1=R,\;v_2=G,\;v_3=R,\;v_4=G,\;v_5=G\} $.

Table \ref{tab:Summary_of_the_postponed_merge-request_validation} reports an overview of the test.

\begin{table}[ht] \centering \caption{Outcome summary of Test 5.} \label{tab:Summary_of_the_postponed_merge-request_validation}
\scriptsize
 \begin{tabular}{ll} 
 \hline 
 Source conflict cluster & $K_1=\{v_1,v_2,v_3\}$ \\  
 Target conflict cluster & $K_2=\{v_4,v_5\}$ \\ 
 Source leader & $v_1$ \\  
 Target leader & $v_4$ \\ 
 Boundary constraint & $c_{24}$ \\  
 Deferred update & $v_2:R\rightarrow G$ \\  
 Initial target-cluster state & 
 \texttt{repairing} \\  
 Initial merge response & 
 \texttt{POSTPONE} \\  
 Postponed-queue size after postponement & 1 \\  
 Target-cluster state before recheck & \texttt{idle} \\  
 Merge response after recheck & \texttt{ACCEPT} \\ 
 Postponed-queue size after committed update & 0 \\  
 Explicit messages & 6 \\ 
 Assignment updated after recheck & Yes \\ \hline \end{tabular} \end{table}

This controlled test confirms that the solver does not trigger immediate merge coordination when a neighbouring conflict cluster is busy. Instead, it queues the request, keeps the current assignment, rechecks the item once the neighbouring cluster is idle, commits the update only after approval, and then clears the queue. The test validates queue handling at the component level; its integration into the full solver loop is tested separately.

\subsection{Impact of the proposed ``memory-guided'' and ``dynamic conflict clustering'' modules}
\label{subsec:diagnostic_scalability_initial_prototype}

We evaluate the influence of the proposed memory-guided restart module on algorithmic scalability and performance by comparing DCC-Incomplete with DCC-Incomplete-random across all benchmark instances for which a planted satisfying assignment is known. Under these conditions, unsuccessful executions of the algorithm do not indicate that the instances are unsatisfiable; rather, they signify that the solver failed to locate a satisfying assignment within the prescribed computational limits.
 
Table~\ref{tab:controlled_family_diagnostic_scalability} reports the numerical results for DCC-Incomplete-random, disaggregated by problem family and instance size, in order to assess the scalability of the algorithm as the number of variables increases within each family. The reported numerical entries correspond to median values computed over 6 runs (3 graph seeds $\times$ 2 initial-assignment seeds). DCC-Incomplete-random solves small instances, but performance declines as graph size and constraint density grow. It solves most 20-variable instances, including all M-family cases. At 50 variables, random restarts fail on sparse family problems and cannot solve any M or D instances. At 100 variables, it solves only one S instance, with all M and D instances unsolved. The XD stress test confirms this trend. All 20-variable instances were solved, but the success rate fell to 50.00\% for 50 variables and 0.00\% for 100 variables. Runtime, iterations, messages, restarts, and stagnation events also increased in the harder settings.

\begin{table}[ht!]
\caption{Numerical results for DCC-Incomplete-random across 6 runs per configuration. SR and $N_{SC}$ report solved-case summaries; the remaining numerical metrics are
reported as medians.}
\label{tab:controlled_family_diagnostic_scalability}
\centering
\begin{adjustbox}{max width=0.9\textwidth}
\begin{tabular}{lllllllllll}
\hline
\textbf{F} & $\boldsymbol{N_V}$ & $\boldsymbol{N_C}$ & \textbf{SR} & $\boldsymbol{\rho}$ & \textbf{Iter.} & $\boldsymbol{N_M}$ & $\boldsymbol{N_{M/V}}$ & $\boldsymbol{N_{M/C}}$ & $\boldsymbol{N_R}$ & $\boldsymbol{FC}$ \\
\hline

S  & 20  & 40  & 66.67\%  & 0.061 & 97.5  & 667.5  & 33.375 & 16.688 & 31.5 & 0.0 \\
S  & 50  & 100 & 50.00\%  & 0.350 & 141.5 & 1206.0 & 24.120 & 12.060 & 39.0 & 0.5 \\
S  & 100 & 200 & 16.67\%  & 1.535 & 150.0 & 1452.0 & 14.520 & 7.260  & 45.5 & 8.0 \\

\hline

M  & 20  & 46  & 100.00\% & 0.058 & 55.0  & 455.5  & 22.775 & 9.902  & 14.5 & 0.0 \\
M  & 50  & 115 & 0.00\%   & 0.512 & 150.0 & 1539.5 & 30.790 & 13.387 & 42.0 & 7.0 \\
M  & 100 & 230 & 0.00\%   & 2.040 & 150.0 & 1699.5 & 16.995 & 7.389  & 44.0 & 11.0 \\

\hline

D  & 20  & 50  & 66.67\%  & 0.120 & 115.0 & 959.0  & 47.950 & 19.180 & 32.0 & 0.0 \\
D  & 50  & 125 & 0.00\%   & 0.624 & 150.0 & 1933.0 & 38.660 & 15.464 & 39.0 & 9.5 \\
D  & 100 & 250 & 0.00\%   & 2.624 & 150.0 & 2320.5 & 23.205 & 9.282  & 36.0 & 18.5 \\

\hline

XD & 20  & 54  & 100.00\% & 0.141 & 113.0 & 1213.0 & 60.650 & 22.463 & 31.5 & 0.0 \\
XD & 50  & 135 & 50.00\%  & 1.374 & 273.0 & 3492.0 & 69.840 & 25.867 & 73.0 & 3.5 \\
XD & 100 & 270 & 0.00\%   & 5.381 & 300.0 & 5417.5 & 54.175 & 20.065 & 79.0 & 25.0 \\
\hline
\end{tabular}
\end{adjustbox}
\end{table}

Table~\ref{tab:dcc_memory_guided_numerical_results} reports the numerical results for DCC-Incomplete. As in the previews table, these are median values across 6 runs displayed per problem family and instance size. Clearly, this variant with the memory-guided restart module performs better, solving all 20-variable instances in the S, D, and XD families and all 50-variable instances in the S family. It also yields better results on the 50-variable M, D, and XD cases, though the 100-variable M, D, and XD cases remain unsolved within the stopping limits. However, it should be noted that in these unsolved cases this variant usually reduced the median number of iterations, messages per constraint, and final violations. This observation indicates that the algorithm is operating more efficiently and could therefore be executed for a longer duration (i.e., with a higher maximum number of iterations) in order to potentially achieve improved results.

\begin{table}[ht!]
\caption{Numerical results for DCC-Incomplete across 6 runs per configuration. SR and $N_{SC}$ report solved-case summaries; the remaining numerical metrics are
reported as medians.}
\label{tab:dcc_memory_guided_numerical_results}
\centering
\begin{adjustbox}{max width=0.9\textwidth}
\begin{tabular}{lllllllllll}
\hline
\textbf{F} & $\boldsymbol{N_V}$ & $\boldsymbol{N_C}$ & \textbf{SR} & $\boldsymbol{\rho}$ & \textbf{Iter.} & $\boldsymbol{N_M}$ & $\boldsymbol{N_{M/V}}$ & $\boldsymbol{N_{M/C}}$ & $\boldsymbol{N_R}$ & $\boldsymbol{FC}$ \\
\hline
S  & 20  & 40  & 100.00\% & 0.030 & 33.0  & 202.0  & 10.100 & 5.050  & 11.0  & 0.0  \\
S  & 50  & 100 & 100.00\% & 0.128 & 42.5  & 334.0  & 6.680  & 3.340  & 15.0  & 0.0  \\
S  & 100 & 200 & 50.00\%  & 1.353 & 145.5 & 1078.0 & 10.780 & 5.390  & 64.0  & 0.5  \\

\hline

M  & 20  & 46  & 83.33\%  & 0.051 & 56.5  & 334.0  & 16.700 & 7.261  & 20.5  & 0.0  \\
M  & 50  & 115 & 66.67\%  & 0.434 & 114.5 & 1008.0 & 20.160 & 8.765  & 60.5  & 0.0  \\
M  & 100 & 230 & 0.00\%   & 1.871 & 143.5 & 1536.5 & 15.365 & 6.680  & 75.0  & 10.0 \\

\hline

D  & 20  & 50  & 100.00\% & 0.041 & 24.5  & 229.0  & 11.450 & 4.580  & 8.0   & 0.0  \\
D  & 50  & 125 & 50.00\%  & 0.539 & 112.0 & 1245.0 & 24.900 & 9.960  & 60.0  & 1.5  \\
D  & 100 & 250 & 0.00\%   & 2.368 & 148.0 & 1922.0 & 19.220 & 7.688  & 75.0  & 9.5  \\

\hline

XD & 20  & 54  & 100.00\% & 0.020 & 19.0  & 199.5  & 9.975  & 3.694  & 4.0   & 0.0  \\
XD & 50  & 135 & 83.33\%  & 0.650 & 130.5 & 1147.5 & 22.950 & 8.500  & 56.0  & 0.0  \\
XD & 100 & 270 & 0.00\%   & 4.728 & 269.5 & 3698.5 & 36.985 & 13.698 & 150.0 & 10.0 \\
\hline
\end{tabular}
\end{adjustbox}
\end{table}

Compared to the random restart variant, DCC-Incomplete performs better in terms of number of solved problems (Table~\ref{tab:family_size_comparison_random_memory_restart}), and generally reduces runtime, iterations, communication and final violations, indicating that memory-guided restart improves the repair trajectory in difficult cases (Table \ref{tab:summary_comparison_random_memory_restart}). The largest gain was in the 50-variable cases: the random-restart version solved no M or D instances, while the memory-guided version solved four M and three D instances. Under the XD stress-test family, both variants solved all 20-variable cases, but memory-guided restart increased the 50-variable success rate from 50.00\% to 83.33\%.

\begin{table}
\centering
\caption{Comparison of DCC-Incomplete-random and DCC-Incomplete with respect to the number of successfully solved runs per instance family and problem size.}
\label{tab:family_size_comparison_random_memory_restart}
\begin{adjustbox}{max width=0.5\textwidth}
\centering
\begin{tabular}{llcc}
\toprule
\textbf{F} &
$\boldsymbol{n}$ &
\textbf{DCC-Incomplete-random} &
\textbf{DCC-Incomplete} \\
\midrule
S & 20 &  4/6 & \textbf{6/6} \\
S & 50 &  3/6 & \textbf{6/6} \\
S & 100 &  1/6 & \textbf{3/6} \\
M & 20 &  \textbf{6/6} & 5/6 \\
M & 50 &  0/6 & \textbf{4/6} \\
M & 100 &  0/6 & 0/6 \\
D & 20 &  4/6 & \textbf{6/6} \\
D & 50 &  0/6 & \textbf{3/}6 \\
D & 100 &  0/6 & 0/6 \\
XD & 20 &  6/6 & 6/6 \\
XD & 50 &  3/6 & \textbf{5/6} \\
XD & 100 &  0/6 & 0/6 \\
\hline
\multicolumn{2}{c}{Total} & 27/72 & \textbf{44/72}\\
\hline
\end{tabular}
\end{adjustbox}
\end{table}

Table~\ref{tab:summary_comparison_random_memory_restart} presents a comparative analysis based on the results obtained from all 72 benchmark runs within our experimental setup. Note that the median number of local restarts $N_R$ was nearly the same for the two variants, so the improvement recorded such mechanism made restarts more effective by reducing repeated local states and choosing better restart candidates, lowering the mean final violations from 7.81 to 2.81.

\begin{table}[ht!]
\centering
\caption{Comparison between DCC-Incomplete-random and DCC-Incomplete over the complete set of 72 benchmark cases. SR reports the success rate (percentage), $N_{SC}$ reports the number of solved cases (count), $\overline{FC}$ reports the mean final cost, and all remaining numerical metrics are reported as medians.}
\label{tab:summary_comparison_random_memory_restart}
\small

\begin{tabular}{lcc}
\toprule
\textbf{Metric} & \textbf{Random restart} & \textbf{Memory-guided restart} \\
\midrule
SR & 37.50\% & 61.11\% \\
$\rho$ & 0.565 & 0.367 \\
Iter. & 150.0 & 98.0 \\
$N_M$ & 1,481.0 & 765.0 \\
$N_{M/V}$ & 28.635 & 14.905 \\
$N_{M/C}$ & 12.493 & 6.503 \\
$\overline{FC}$ & 7.81 & 2.81 \\
$N_R$ & 40.0 & 41.0 \\
\bottomrule
\end{tabular}
\end{table}

To provide additional evidence supporting the use of the proposed memory-guided restart mechanisms, we designed two experiments to verify that simply increasing the maximum number of iterations and the maximum number of allowed restarts (denoted as $N_R^{MAX}$) in DCC-Incomplete-random would not yield comparable results (i.e., would not produce fewer failures than those reported in Table~\ref{tab:family_size_comparison_random_memory_restart}). 

We selected the $12$ challenging unsolved instances from the
harder medium, dense, and extreme-dense configurations with 50 and 100 variables, and attempted to solve them running DCC-Incomplete-random with hyperparameter setting \(T_{\max}=1500\) and \(N_R^{\max}=200\)\footnote{The algorithms are stopped if they have solved the problem or when either \(T_{\text{max}}\) or \(N_R^{\text{max}}\) is reached.} (Experiment 1, $m=12$). As reported in Table~\ref{tab:stopping_limit_sensitivity}, six instances were solved (i.e., a SR of 50\%). Consequently, we executed the algorithm again on the six remaining unsolved instances, this time setting \(T_{\max}=3000\) and \(N_R^{\max}=400\) (Experiment 2, $m=6$). This time, only 1 instance was solved (i.e., SR dropped to $16.67$\%). Hence, relaxing (twice) the
stopping thresholds solved only 7 additional difficult cases, while 5 cases still remained unsolved and at the cost of a substantial increase in search and communication overhead and elevated number of restarts. A cursory examination of Table \ref{tab:stopping_limit_sensitivity} indicates that merely increasing the stopping limits can resolve a modest subset of additional instances; however, this approach is neither efficient nor sustainable as a general strategy. This observation underscores the importance of the memory-guided restart mechanism, which yields superior results while requiring a substantially lower computational budget to execute the algorithm.

\begin{table}[ht!]
\centering
\caption{Results of the extended stopping-budget experiments for DCC-Incomplete-random. SR and $N_{SC}$ report
solved-case summaries within each subset; $\overline{FC}$ reports mean final cost; the remaining numerical metrics are reported as medians.}
\label{tab:stopping_limit_sensitivity}
\small
\begin{tabular}{lcc}
\hline
\textbf{Metric} & \textbf{Experiment 1} & \textbf{Experiment 2} \\
\hline
SR & 50.0\% & 16.7\% \\
Iter. & 599.5 & 1217.0 \\
$N_R$ & 192.0 & 400.0 \\
$N_{M/V}$ & 102.27 & 169.99 \\
$N_{M/C}$ & 40.95 & 66.98 \\
$\overline{FC}$ & 8.25 & 10.67 \\
$N_{SC}$ & 6/12 & 1/6 \\
\hline
\end{tabular}%

\end{table}

Finally, we show how the dynamic conflict clustering module reduces communication overhead by using agents more efficiently. In this framework, repair activity is concentrated on the agents currently involved in conflicts, rather than having all agents continuously and inefficiently work on the global problem. Using the same experimental setup as for the two restart modules, we define four metrics for a generic iteration $t\in\{1,2,\dots,T_{max}\}$: (1) mean active agent index(\(\mu^t_{AAI}\)), the proportion of active agents in the current runtime state,
computed as the number of active agents normalised by the total number of agents; (2) \emph{mean conflict cluster size} ($\mu_{CC}^t$), the average number of agents/variables per conflict cluster; (3) \emph{maximum observed cluster size} ($M^t_{CC}$), the size of the largest conflict cluster; and (4) \emph{mean number of local restarts} ($\mu_{LR}$), the average number of local restarts triggered during recovery from stagnation. At the end of each run, the values recorded over the executed iterations \(t\) are aggregated to obtain a single summary value per run: \(\mu^t_{AAI}\) and \(\mu^t_{CC}\) are averaged over the executed iterations, while \(M^t_{CC}\)
records the maximum ConflictCluster size observed during the run. The local-restart
count is recorded per run. The obtained run-level values are then aggregated
across runs for each family-size configuration. In Table \ref{tab:conflict cluster_locality_restart_characteristics}, \(\mu^t_{AAI}\), \(\mu^t_{CC}\), and \(\mu_{LR}\) report mean values across runs, while \(M^t_{CC}\) reports the maximum observed ConflictCluster size.

\begin{table}[ht]
\centering
\caption{Numerical results showing the impact of dynamic conflict clustering. The metrics \(\mu^t_{AAI}\), \(\mu^t_{CC}\), and \(\mu_{LR}\) report mean values across runs for each family-size configuration, while \(M^t_{CC}\) reports the maximum observed ConflictCluster size.}
\label{tab:conflict cluster_locality_restart_characteristics}
\small
\begin{tabular}{llllll}
\toprule
\textbf{F} &
$\boldsymbol{n}$ &
$\boldsymbol{\mu^t_{AAI}}$&
$\boldsymbol{\mu_{cc}^t}$&
$\boldsymbol{M^t_{CC}}$ &
$\boldsymbol{\mu_{LR}}$ \\
\midrule

S  & 20  & 0.270 & 2.89 & 11 & 28.8 \\
M  & 20  & 0.322 & 3.45 & 9  & 14.7 \\
D  & 20  & 0.394 & 3.83 & 11 & 28.2 \\
XD  & 20 & 0.477 & 4.21 & 13 & 32.2 \\

\midrule

S  & 50  & 0.235 & 3.52 & 19 & 29.5 \\
M  & 50  & 0.299 & 3.93 & 20 & 43.7 \\
D  & 50  & 0.357 & 4.58 & 21 & 40.8 \\
XD  & 50 & 0.367 & 4.44 & 22 & 65.7 \\

\midrule

S & 100  & 0.283 & 3.48 & 29 & 43.0 \\
M & 100  & 0.357 & 3.54 & 33 & 43.7 \\
D & 100  & 0.439 & 4.34 & 38 & 36.2 \\
XD & 100 & 0.429 & 4.70 & 41 & 80.2 \\

\bottomrule
\end{tabular}
\end{table}

The locality measurements summarised in Table \ref{tab:conflict cluster_locality_restart_characteristics} show that the mean conflict cluster size remained below five variables across all tested configurations, including the XD stress-test family. This supports the intended locality behaviour of DCC-Incomplete: repair is usually restricted to small currently conflicted regions rather than repeatedly activating the full graph.

The active-variable ratio increased with density, as expected, but remained below one in all cases. Therefore, the main limitation of the initial random-restart prototype was not uncontrolled cluster expansion. Instead, difficult cases were 
mainly associated with repeated local restart activity, which motivated the memory-guided restart refinement.

\subsection{Comparison with the state-of-the-art}
\label{subsec:performance-evaluation-external-methods}

\newcolumntype{L}[1]{>{\raggedright\arraybackslash}p{#1}}
\newcolumntype{C}[1]{>{\centering\arraybackslash}p{#1}}
\newcolumntype{R}[1]{>{\raggedleft\arraybackslash}p{#1}}

\newcommand{\MetricTableMetricCol}{3.05cm}
\newcommand{\MetricTableMethodCol}{1.28cm}
\newcommand{\FamilyCol}{0.55cm}
\newcommand{\VarsCol}{0.70cm}
\newcommand{\MetricCol}{2.65cm}
\newcommand{\DCCMemCol}{1.18cm}
\newcommand{\DCCRandCol}{1.18cm}
\newcommand{\DSACol}{1.02cm}
\newcommand{\GDBACol}{1.12cm}
\newcommand{\MGMCol}{1.12cm}

 We emphasise that GDBA does not support a cycle-based stopping mechanism and instead relies exclusively on a strict timeout criterion (10 seconds in our experiments\footnote{The reported time may slightly exceed this value because it includes collecting results after the algorithm stops.}). For this algorithm, the maximum iteration count is therefore not applicable. Although this is not an issue for many instances -- since the algorithm often terminates successfully before completing 150 iterations -- it may lead to slightly unfair comparisons in certain cases. Furthermore, the DCC repair iterations are not directly comparable to the cycles counted by pyDCOP. For this reason, the Iter. metric is reported for transparency, but we do not base our analysis on it. Instead, we principally compare the algorithms using success rate, final cost, runtime, and communication cost metrics. 

Moreover, as previously stated in Section \ref{subsec:experimental_protocol}, we explicitly track and count each message exchanged during the execution of our algorithms, whereas pyDCOP’s internal message handling can prevent an exact message count. Consequently, we do not focus on the absolute number of messages exchanged; instead, we consider their normalized values per constraint, which provides a valid basis for comparing the algorithms.  

Note that DCC-Incomplete is designed for DCSPs, whereas the comparison algorithms are implemented within a DCOP framework. This distinction should not cause confusion, as any DCSP can be modeled as a special case of a DCOP, thereby ensuring the validity of our comparative analysis.

Our results (Table~\ref{tab:performance-overall}) show that DCC-Incomplete achieves the highest success rate of 61.11\%, followed by DCC-Incomplete-random with 37.50\% and DSA with 35.56\%. GDBA and MGM achieve success rates of 3.61\% and 0.83\%, respectively. DCC-Incomplete also achieves the lowest reported final cost and substantially lower communication overhead than external baselines.

\begin{table}[ht!]
    \centering
    \caption{Comparison between the five algorithms over the complete set of 72 benchmark cases. Baseline values are averaged over 10 executions; SR is computed over the 72 cases, and other metrics use median summaries.}
    \label{tab:performance-overall}
    \begin{adjustbox}{max width=0.99\textwidth}
    \begin{tabular}{llllll}
        \toprule
        \textbf{Metric} & \textbf{DCC-Incomplete} & \textbf{DCC-Incomplete-random} & \textbf{DSA} & \textbf{GDBA} & \textbf{MGM} \\
        \midrule
        SR & 61.11\% & 37.50\% & 35.56\% & 3.61\% & 0.83\% \\
        $\rho$ & 0.367 & 0.565 & 2.395 & 10.098 & 8.047 \\
        Iter & 98.0 & 150.0 & 150.0 & 58.55 & 150.0 \\
        ${N_{M}}$ & 765.0 & 1481.0 & 36440.0 & 27741.5 & 71960.0 \\
        ${N_{M/V}}$ & 14.905 & 28.635 & 728.800 & 563.070 & 1439.200 \\
        ${N_{M/C}}$ & 6.503 & 12.493 & 303.670 & 234.157 & 599.670 \\
        FC & 0.0 & 5.0 & 1.20 & 7.15 & 9.10 \\
        \bottomrule
    \end{tabular}
    \end{adjustbox}
\end{table}

With respect to communication complexity, there is a substantial difference between the proposed algorithms and the baseline methods. Specifically, DCC-Incomplete returns a median of 14.905 messages per variable and 6.503 messages per constraint, whereas DSA, GDBA, and MGM yield substantially higher values. These results further substantiate our claim that the proposed algorithm continue reduced communication overhead.

To further analyse performance across problem families and sizes, we provide results at the level of individual problem instances, organised by family: S (Table \ref{tab:Comparison_between_five_algorithms-family_S}), M (Table \ref{tab:Comparison_between_five_algorithms-family_M}), D (Table \ref{tab:Comparison_between_five_algorithms-family_D}), and XD (Table \ref{tab:fivealgorithmsonxd}).

\begin{table}[ht!]
    \centering
    \caption{Comparison of five algorithms on family S instances.}
    \label{tab:Comparison_between_five_algorithms-family_S}
    \begin{adjustbox}{max width=\textwidth}
 \begingroup
    \renewcommand{\arraystretch}{0.9}
    \begin{tabular}{lllllll}
        \toprule
        $n$ & Metric & DCC-Incomplete & DCC-Incomplete-random & DSA & GDBA & MGM \\
        \midrule
        20 & SR & 100.00 & 66.67 & 76.67 & 10.00 & 3.33 \\
           & $\rho$  & 0.030 & 0.061 & 0.798 & 10.032 & 2.824 \\
           & Iter & 33.0 & 97.5 & 150.0 & 173.50 & 150.0 \\
           & ${N_{M}}$ & 202.0 & 667.5 & 12160.0 & 27771.30 & 24000.0 \\
           & ${N_{M/V}}$ & 10.100 & 33.375 & 608.000 & 1388.565 & 1200.000 \\
           & ${N_{M/C}}$ & 5.050 & 16.688 & 304.000 & 694.283 & 600.000 \\
           & FC & 0.0 & 0.0 & 0.15 & 2.10 & 2.30 \\
        \midrule
        50 & SR & 100.00 & 50.00 & 51.67 & 0.00 & 0.00 \\
           & $\rho$ & 0.128 & 0.350 & 2.047 & 10.115 & 7.080 \\
           & Iter & 42.5 & 141.5 & 150.0 & 69.55 & 150.0 \\
           & ${N_{M}}$ & 334.0 & 1206.0 & 30400.0 & 27822.80 & 60000.0 \\
           & ${N_{M/V}}$ & 6.680 & 24.120 & 608.000 & 556.456 & 1200.000 \\
           & ${N_{M/C}}$ & 3.340 & 12.060 & 304.000 & 278.228 & 600.000 \\
           & FC & 0.0 & 0.5 & 0.50 & 4.70 & 5.80 \\
        \midrule
        100 & SR & 50.00 & 16.67 & 6.67 & 0.00 & 0.00 \\
            & $\rho$ & 1.353 & 1.535 & 4.238 & 10.178 & 14.443 \\
            & Iter & 145.5 & 150.0 & 150.0 & 28.20 & 150.0 \\
            & ${N_{M}}$ & 1078.0 & 1452.0 & 60800.0 & 22416.80 & 120000.0 \\
            & ${N_{M/V}}$ & 10.780 & 14.520 & 608.000 & 224.168 & 1200.000 \\
            & ${N_{M/C}}$ & 5.390 & 7.260 & 304.000 & 112.084 & 600.000 \\
            & FC & 0.5 & 8.0 & 2.35 & 10.40 & 12.65 \\
        \bottomrule
    \end{tabular}
    \endgroup
    \end{adjustbox}
\end{table}

\begin{table}[ht!]
    \centering
    
    \caption{Comparison between the five algorithms for family M instances.}
    \label{tab:Comparison_between_five_algorithms-family_M}
    \begin{adjustbox}{max width=\textwidth}
    \begingroup
    \renewcommand{\arraystretch}{0.9}
    \begin{tabular}{lllllll}
        \toprule
        $n$ & Metric & DCC-Incomplete & DCC-Incomplete-random & DSA & GDBA & MGM \\
        \midrule
        20 & SR & 83.33 & 100.00 & 55.00 & 5.00 & 3.33 \\
           & $\rho$ & 0.051 & 0.058 & 0.898 & 10.034 & 3.124 \\
           & Iter & 56.5 & 55.0 & 150.0 & 154.30 & 150.0 \\
           & ${N_{M}}$  & 334.0 & 455.5 & 13972.0 & 28425.15 & 27588.0 \\
           & ${N_{M/V}}$  & 16.700 & 22.775 & 698.600 & 1421.258 & 1379.400 \\
           & ${N_{M/C}}$  & 7.261 & 9.902 & 303.739 & 617.938 & 599.739 \\
           & FC & 0.0 & 0.0 & 0.45 & 2.95 & 3.10 \\
        \midrule
        50 & SR & 66.67 & 0.00 & 10.00 & 0.00 & 0.00 \\
           & $\rho$ & 0.434 & 0.512 & 2.338 & 10.102 & 7.817 \\
           & Iter & 114.5 & 150.0 & 150.0 & 60.90 & 150.0 \\
           & ${N_{M}}$ & 1008.0 & 1539.5 & 34930.0 & 27997.60 & 68970.0 \\
           & ${N_{M/V}}$ & 20.160 & 30.790 & 698.600 & 559.952 & 1379.400 \\
           & ${N_{M/C}}$ & 8.765 & 13.387 & 303.739 & 243.457 & 599.739 \\
           & FC & 0.0 & 7.0 & 1.75 & 7.05 & 8.15 \\
        \midrule
        100 & SR & 0.00 & 0.00 & 1.67 & 0.00 & 0.00 \\
            & $\rho$ & 1.871 & 2.040 & 4.719 & 10.199 & 15.901 \\
            & Iter & 143.5 & 150.0 & 150.0 & 23.70 & 150.0 \\
            & ${N_{M}}$ & 1536.5 & 1699.5 & 69860.0 & 21819.00 & 137940.0 \\
            & ${N_{M/V}}$ & 15.365 & 16.995 & 698.600 & 218.190 & 1379.400 \\
            & ${N_{M/C}}$ & 6.680 & 7.389 & 303.739 & 94.865 & 599.739 \\
            & FC & 10.0 & 11.0 & 3.70 & 13.50 & 16.45 \\
        \bottomrule
    \end{tabular}
    \endgroup
    \end{adjustbox}
\end{table}

\begin{table}[ht!]
    \centering
    \caption{Comparison between the five algorithms for family D instances. }
    \label{tab:Comparison_between_five_algorithms-family_D}
    \begin{adjustbox}{max width=\textwidth}
     \begingroup
    \renewcommand{\arraystretch}{0.9}
    \begin{tabular}{lllllll}
        \toprule
        $n$ & Metric & DCC-Incomplete & DCC-Incomplete-random & DSA & GDBA & MGM \\
        \midrule
        20 & SR & 100.00 & 66.67 & 66.67 & 8.33 & 0.00 \\
           & $\rho$ & 0.041 & 0.120 & 0.999 & 10.041 & 3.384 \\
           & Iter & 24.5 & 115.0 & 150.0 & 143.30 & 150.0 \\
           & ${N_{M}}$ & 229.0 & 959.0 & 15180.0 & 28673.40 & 29980.0 \\
           & ${N_{M/V}}$ & 11.450 & 47.950 & 759.000 & 1433.670 & 1499.000 \\
           & ${N_{M/C}}$ & 4.580 & 19.180 & 303.600 & 573.468 & 599.600 \\
           & FC & 0.0 & 0.0 & 0.30 & 3.55 & 4.45 \\
        \midrule
        50 & SR & 50.00 & 0.00 & 23.33 & 0.00 & 0.00 \\
           & $\rho$ & 0.539 & 0.624 & 2.448 & 10.114 & 8.210 \\
           & Iter & 112.0 & 150.0 & 150.0 & 56.35 & 150.0 \\
           & ${N_{M}}$ & 1245.0 & 1933.0 & 37950.0 & 28200.25 & 74950.0 \\
           & ${N_{M/V}}$ & 24.900 & 38.660 & 759.000 & 564.005 & 1499.000 \\
           & ${N_{M/C}}$ & 9.960 & 15.464 & 303.600 & 225.602 & 599.600 \\
           & FC & 1.5 & 9.5 & 1.70 & 8.70 & 10.25 \\
        \midrule
        100 & SR & 0.00 & 0.00 & 1.67 & 0.00 & 0.00 \\
            & $\rho$ & 2.368 & 2.624 & 5.038 & 10.211 & 16.962 \\
            & Iter & 148.0 & 150.0 & 150.0 & 21.65 & 150.0 \\
            & ${N_{M}}$ & 1922.0 & 2320.5 & 75900.0 & 21487.40 & 149900.0 \\
            & ${N_{M/V}}$ & 19.220 & 23.205 & 759.000 & 214.874 & 1499.000 \\
            & ${N_{M/C}}$ & 7.688 & 9.282 & 303.600 & 85.950 & 599.600 \\
            & FC & 9.5 & 18.5 & 5.70 & 16.70 & 19.65 \\
        \bottomrule
    \end{tabular}
    \endgroup
    \end{adjustbox}
\end{table}

\begin{table}[ht!]
    \centering
    \caption{Comparison between the five algorithms for family XD instances.}
    \label{tab:fivealgorithmsonxd}
    \begin{adjustbox}{max width=\textwidth}
     \begingroup
    \renewcommand{\arraystretch}{0.9}
    \begin{tabular}{lllllll}
        \toprule
        $n$ & Metric & DCC-Incomplete & DCC-Incomplete-random & DSA & GDBA & MGM \\
        \midrule
        20 & SR & 100.00 & 100.00 & 90.00 & 20.00 & 3.33 \\
           & $\rho$ & 0.020 & 0.141 & 1.043 & 10.032 & 3.576 \\
           & Iter & 19.0 & 113.0 & 150.0 & 134.80 & 150.0 \\
           & ${N_{M}}$ & 199.5 & 1213.0 & 16388.0 & 29148.95 & 32372.0 \\
           & ${N_{M/V}}$ & 9.975 & 60.650 & 819.400 & 1457.448 & 1618.600 \\
           & ${N_{M/C}}$ & 3.694 & 22.463 & 303.481 & 539.795 & 599.481 \\
           & FC & 0.0 & 0.0 & 0.00 & 3.00 & 3.90 \\
        \midrule
        50 & SR & 83.33 & 50.00 & 38.33 & 0.00 & 0.00 \\
           & $\rho$ & 0.650 & 1.374 & 2.608 & 10.142 & 8.700 \\
           & Iter & 130.5 & 273.0 & 150.0 & 52.40 & 150.0 \\
           & ${N_{M}}$ & 1147.5 & 3492.0 & 40970.0 & 28290.00 & 80930.0 \\
           & ${N_{M/V}}$ & 22.950 & 69.840 & 819.400 & 565.800 & 1618.600 \\
           & ${N_{M/C}}$ & 8.500 & 25.867 & 303.481 & 209.556 & 599.481 \\
           & FC & 0.0 & 3.5 & 1.50 & 10.05 & 11.95 \\
        \midrule
        100 & SR & 0.00 & 0.00 & 5.00 & 0.00 & 0.00 \\
            & $\rho$ & 4.728 & 5.381 & 5.396 & 10.279 & 18.181 \\
            & Iter & 269.5 & 300.0 & 150.0 & 19.65 & 150.0 \\
            & ${N_{M}}$ & 3698.5 & 5417.5 & 81940.0 & 21054.25 & 161860.0 \\
            & ${N_{M/V}}$ & 36.985 & 54.175 & 819.400 & 210.543 & 1618.600 \\
            & ${N_{M/C}}$ & 13.698 & 20.065 & 303.481 & 77.979 & 599.481 \\
            & FC & 10.0 & 25.0 & 6.05 & 20.55 & 24.30 \\
        \bottomrule
    \end{tabular}
    \endgroup
    \end{adjustbox}
\end{table}

\begin{figure}[ht!]
    \centering
    \begin{subfigure}[b]{0.48\textwidth}
        \centering
        \includegraphics[width=\linewidth]{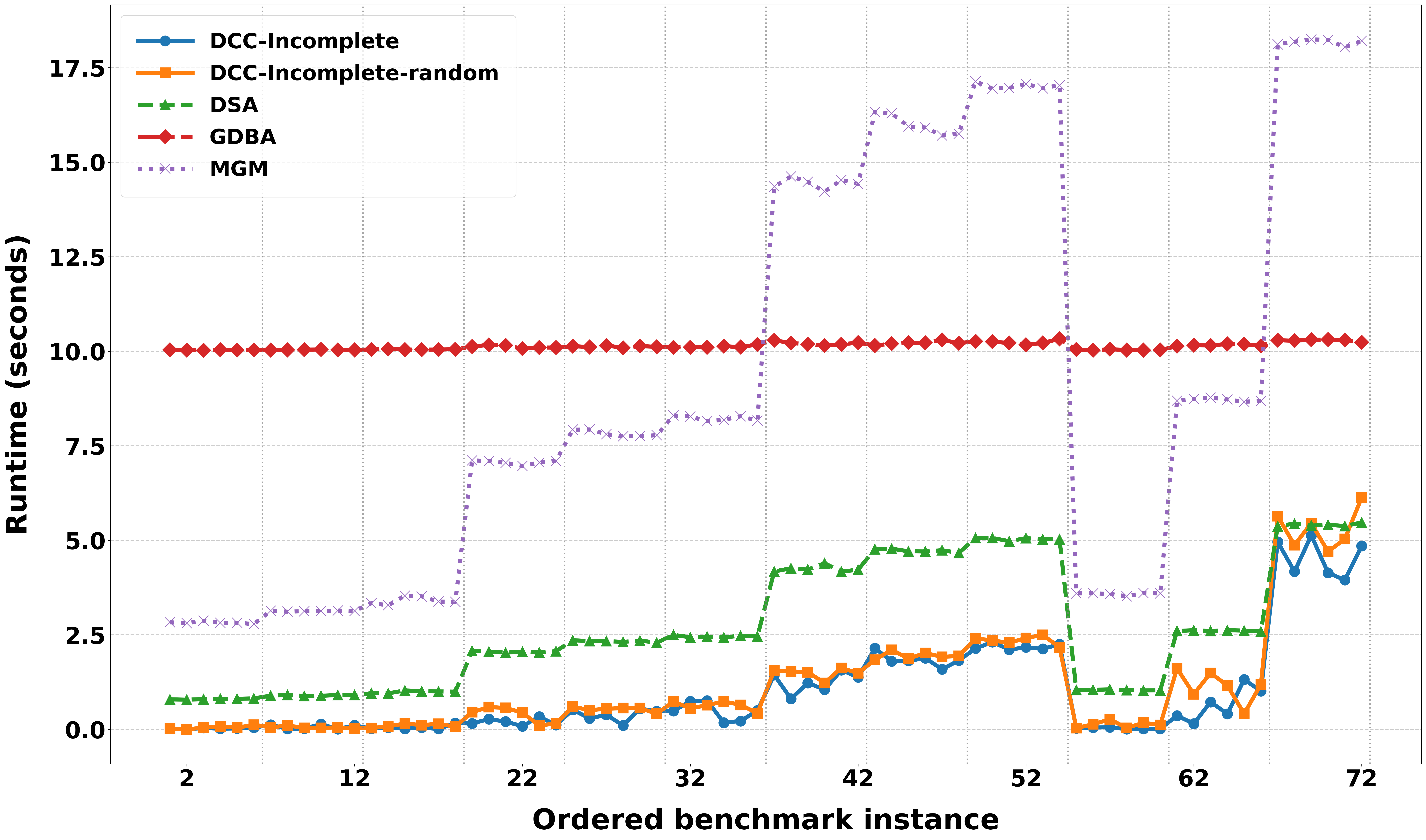} 
        \caption{Runtime across benchmark instances}
        \label{fig:my-3x2-a}
    \end{subfigure}
    \hfill
     \begin{subfigure}[b]{0.48\textwidth}
        \centering
        \includegraphics[width=\linewidth]{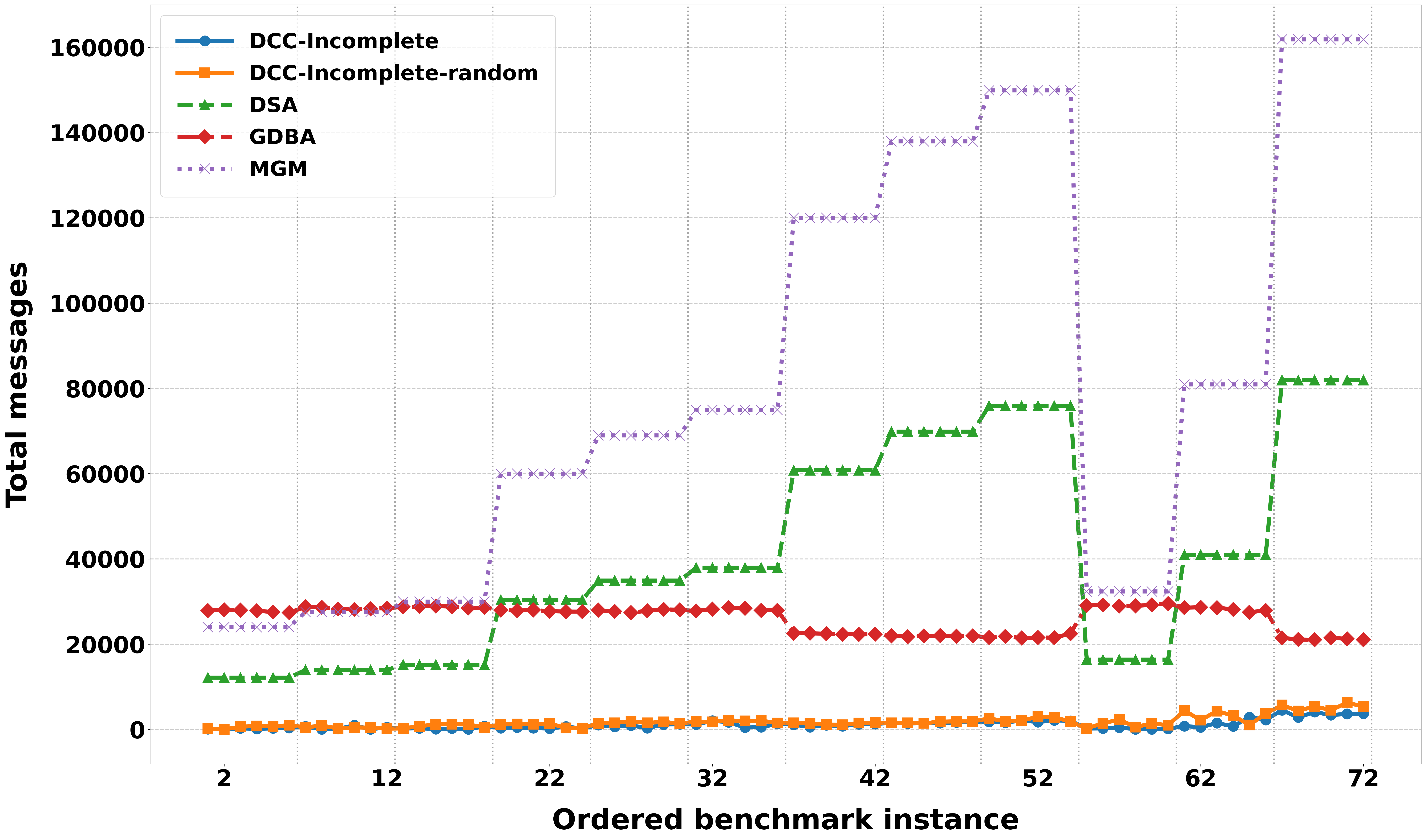}
        \caption{Total messages across benchmark instances}
        \label{fig:}
    \end{subfigure}

    \medskip

    \begin{subfigure}[b]{0.48\textwidth}
        \centering
        \includegraphics[width=\linewidth]{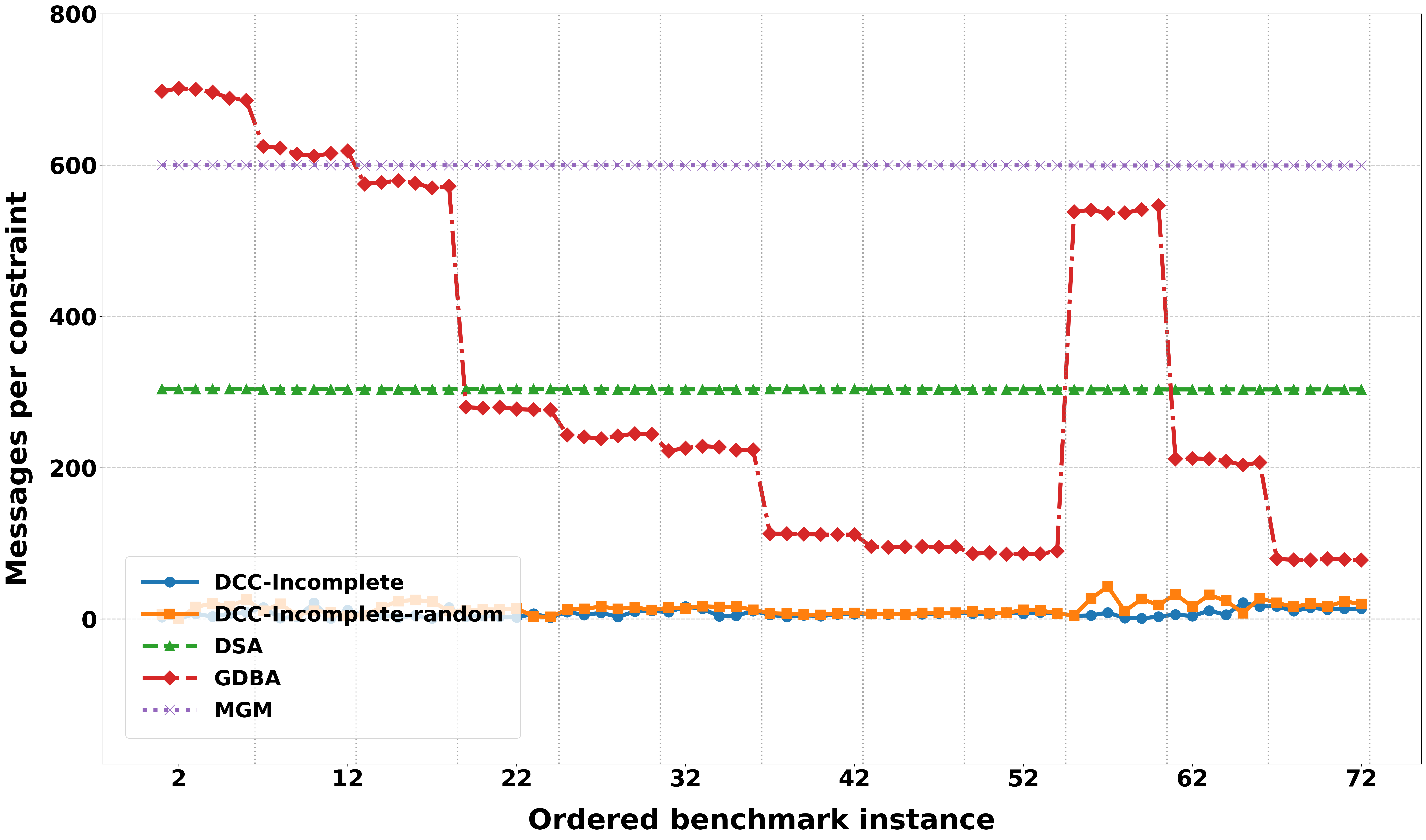} 
        \caption{Messages per constraint across benchmark instances}
        \label{fig:messages_per_constraint_across_benchmark_instances}
    \end{subfigure}
    \hfill
    \begin{subfigure}[b]{0.48\textwidth}
        \includegraphics[width=\linewidth]{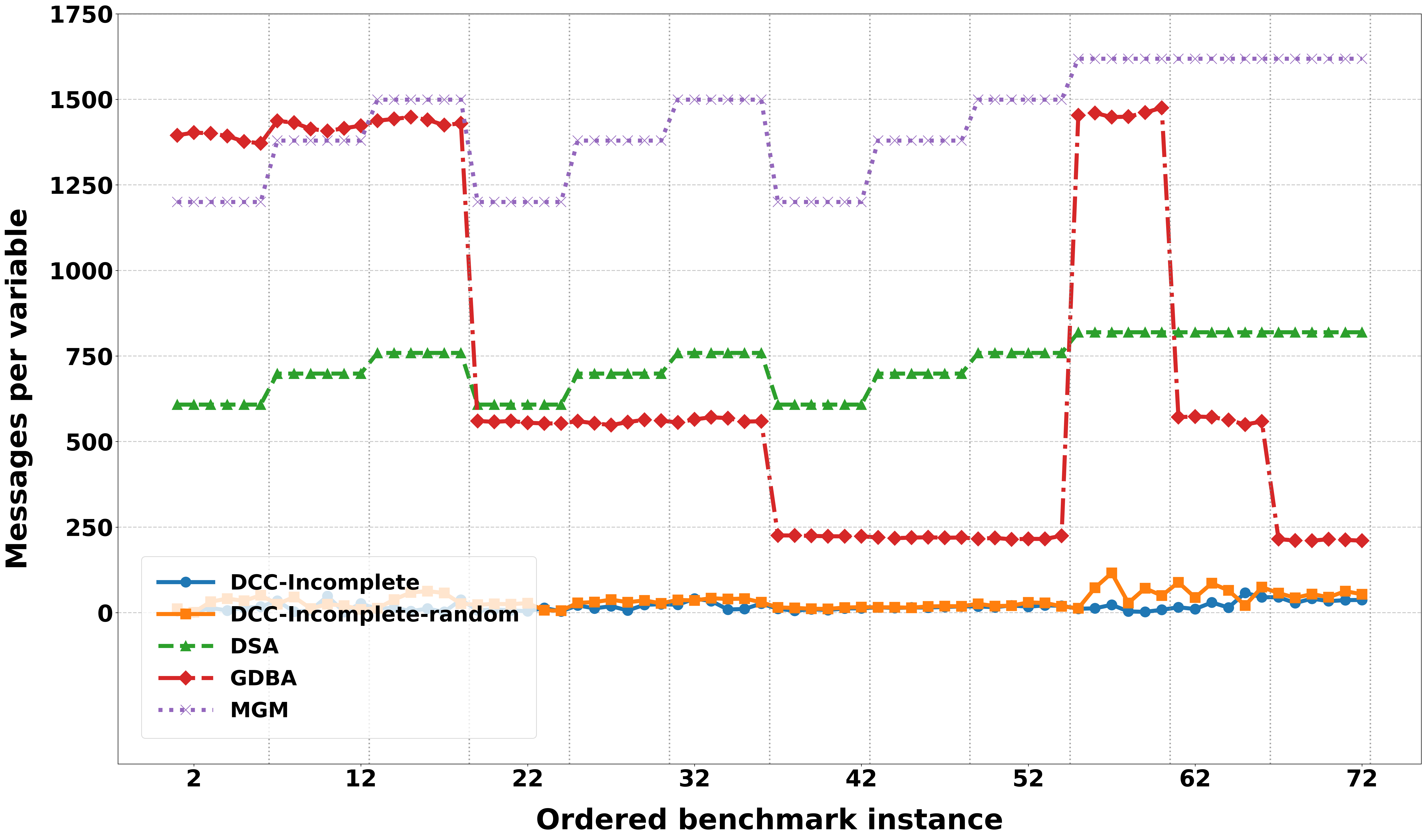} 
        \caption{Messages per variable across benchmark instances}
        \label{fig:messages_per_variable_across_benchmark_instances}
    \end{subfigure}

    \medskip

    \caption{Graphical representation of the nuerical results showing: (a) runtime; (b) total number of messages; (c) messages per constraint; (d) shows messages per variable. } 
    \label{fig:external-performance-dashboard}
\end{figure}

DCC-Incomplete achieves the highest overall success rate and substantially lower communication overhead across the complete benchmark, although some individual family-size configurations show local differences. For sparse instances  (reported in Table \ref{tab:Comparison_between_five_algorithms-family_S}) with 100 variables, DCC-Incomplete achieves a success rate of 50.00\%, compared with 16.67\% for DCC-Incomplete-random and 6.67\% for DSA. In the XD stress-test  (Table \ref{tab:fivealgorithmsonxd}) setting with 50 variables, the corresponding success rates are 83.33\%, 50.00\%, and 38.33\%, respectively. At 100 variables, DSA also achieves a small non-zero success rate of 5.00\% for XD instances. Nevertheless, the proposed methods maintain substantially lower message-passing overhead across the tested problem sizes, as illustrated in Figure~\ref{fig:external-performance-dashboard}.

Overall, our numerical results confirm DCC-Incomplete as a promising repair-based approach for solving DCSPs, featuring a substantially lower communication overhead than established traditional and state-of-the-art algorithms. It is highly competitive, consistently achieving the highest overall success rate and lowest median final cost among the algorithms in this study. Our experimentation verifies its correct functioning and proves that all of its novel modules contribute positively in terms of performance. It is interesting to note, that despite being less effective without memory-guided restarts, the DCC-Incomplete-random variant still outperforms established algorithms such as DSA, GDBA, and MGM in overall success rate, demonstrating that this new framework is highly effective and provides a strong foundation for future research.

\section{Conclusion and Future Work}
\label{sec:conclusion}

Our results demonstrate that the proposed algorithm constitutes an efficient, incomplete, repair-based solver that overcomes two major limitations affecting current distributed constraint-solving approaches, such as: (i) high communication overhead arising from broad or redundant message passing, and (ii) insufficient support for strategic coordination and learning during the repair process.

When evaluated on controlled distributed graph-colouring benchmarks with planted satisfying assignments, the solver demonstrated its efficiency. In particular, its memory-guided variant outperformed the random-restart variant by increasing the success rate from 37.50\% to 61.11\%, reducing the median runtime from 0.565 seconds to 0.367 seconds, and decreasing the median communication overhead from 12.493 to 6.503 messages per constraint. An external comparison further indicated that DCC-Incomplete achieved the strongest overall performance among the evaluated methods, exhibiting a higher success rate and substantially lower
communication overhead than state-of-the-art algorithms such as DSA, MGM, and GDBA on the tested benchmark set.

In this study, we additionally demonstrate that the integration of dynamic conflict-based decomposition with structured coordination mechanisms and lightweight, memory-guided recovery strategies can enhance the performance of incomplete DCSP solving. Furthermore, we show that the proposed solver 
preserves distributed agent-level repair while adding cluster-level organisation 
to reduce unnecessary communication and improve robustness against stagnation.

Future work includes: (1) evaluating the solver on broader benchmarks, such as varied domain sizes, non-binary constraints, heterogeneous agent structures, and real-world distributed constraint problems; (2) conducting a more systematic parameter-sensitivity analysis of the memory-guided restart mechanism (memory length, candidate pool size, scoring weights); (3) performing additional ablation studies to isolate the impact of each component—conflict cluster formation, boundary coordination, merge handling, and memory-guided restart; and (4) extending the incomplete solver to a complete or hybrid variant that combines conflict clustering with stronger search guarantees.

\end{document}